\documentclass[12pt,preprint]{aastex}
\usepackage{color,ulem}

\newcommand\E{{\cal E}}

\shorttitle{Stellar Disruption by Hard SMBHBs} \shortauthors{Chen,
Liu, \& Magorrian}

\begin{document}

\title{Tidal Disruption of Stellar Objects by Hard Supermassive Black Hole
Binaries}

\author{Xian Chen}
\affil{Department of Astronomy, Peking University, 100871 Beijing,
China}
\email{chenx@bac.pku.edu.cn}

\author{F. K. Liu} \affil{Department of Astronomy, Peking University,
100871 Beijing, China} \email{fkliu@bac.pku.edu.cn}

\and

\author{John Magorrian}
\affil{Rudolf Peierls Center for Theoretical Physics, University of
Oxford, 1 Keble Road, Oxford, OX1 3NP, United Kingdom}
\email{magog@thphys.ox.ac.uk}

\begin{abstract}

Supermassive black hole binaries (SMBHBs) are expected by the
hierarchical galaxy formation model in $\Lambda$CDM cosmology. There
is some evidence in the literature for SMBHBs in AGNs, but there are
few observational constraints on the evolution of SMBHBs in inactive
galaxies and gas-poor mergers. On the theoretical front, it is
unclear how long is needed for a SMBHB in a typical galaxy to
coalesce. In this paper we investigate the tidal interaction between
stars and binary BHs and calculate the tidal disruption rates of
stellar objects by the BH components of binary. We derive the
interaction cross sections between SMBHBs and stars from intensive
numerical scattering experiments with particle number $\sim10^7$ and
calculate the tidal disruption rates by both single and binary BHs
for a sample of realistic galaxy models, taking into account the
general relativistic effect and the loss cone refilling because of
two-body interaction. We estimate the frequency of tidal flares for
different types of galaxies using the BH mass function in the
literature. We find that because of the three-body slingshot effect,
the tidal disruption rate in SMBHB system is more than one order of
magnitude smaller than that in single SMBH system. The difference is
more significant in less massive galaxies and does not depend on
detailed stellar dynamical processes. Our calculations suggest that
comparisons of the calculated tidal disruption rates for both single
and binary BHs and the surveys of X-ray or UV flares at galactic
centers could tell us whether most SMBHs in nearby galaxies are
single and whether the SMBHBs formed in gas-poor galaxy mergers
coalesce rapidly.

\end{abstract}


\section{INTRODUCTION}
\label{introduction}

In cold dark matter (CDM) cosmology, galaxies form hierarchically
and present-day galaxies are the products of successive mergers
\citep{kauffmann00,springel05}. Recent observations show that almost
all galaxies harbor supermassive black holes (SMBHs) at their
centers \citep{richstone98,ferrarese05} and the black hole (BH)
masses tightly correlate with the properties of their host galaxies
such as the mass of stellar bulge \citep{magorrian98,haring04}, the
bulge luminosity \citep{mclure02}, and the nuclear stellar velocity
dispersion \citep{ferrarese00,gebhardt00,tremaine02}. The
correlations between BH mass and galaxy properties imply that the
growth of SMBHs and the formation and evolution of galaxies are
closely linked. The correlation is likely induced by galaxy major
mergers (merging of galaxies with comparable mass) in which both
rapid star formation and gas accretion onto SMBHs are triggered and
the feedback from the central active galactic nuclei (AGNs)
regulates the growth of both SMBHs and galaxy bulges. The
coevolution scenario can successfully explain not only the
correlations between SMBHs and their host galaxies but also many of
the observed evolutions of galaxies and AGNs
\citep{springel05,hopkins06,croton06}.

During galaxy mergers part of the gas originally in galactic plane
is driven to galaxy center, triggering starburst and the accretion
of central SMBH \citep{gaskell85,hernquist95}. If both galaxies in
the merging system harbor SMBHs at their centers, a pair of AGNs
could form which is an intriguing object for both observations and
theories. Because of dynamical friction the two SMBHs quickly sink
to the common center of the two merging galaxies, forming a bound,
compact ($\sim10$ pc) supermassive black hole binary
\citep[SMBHB,][]{begelman80}, which would be difficult to resolve
in any but the closest galaxies using current telescopes. As the
separation of the binary shrinks, dynamical friction becomes less
and less efficient and three-body interaction between SMBHB and
the stars passing by becomes more and more important. At
separations $\sim0.1-1$ pc SMBHB becomes hard in the sense that
the binary loses energy and angular momentum mainly through
three-body interaction. The stars passing by the binary, if not
tidally disrupted or swallowed by the BHs, would be expelled by
the ``slingshot effect'' \citep[][Q96 thereafter]{quinlan96} with
velocities comparable to the host galaxy's escape velocity. If the
refilling of the reservoir of stars on orbits that interact with
the SMBHB is inefficient, the evolution of SMBHB slows down and
the evolution timescale may even exceed the Hubble time, with the
SMBHB separation stalling well outside the radii needed for the
final gravitational wave radiation and coalescence stage
\citep[][Y02 hereafter]{yu02}. This seems inconsistent with that
fact that SMBHBs have not yet been observed in nearby galaxies.
Therefore, many stellar-dynamical mechanisms are proposed to boost
the hardening rate of SMBHBs in gas-poor systems, including
non-equilibrium stellar distribution \citep{milosavljevic03},
Brownian motion of SMBHB \citep{chatterjee03}, and non-spherical
(axisymmetric, triaxial, or irregular) stellar distribution
\citep[Y02;][]{merritt04,berczik06}. However, the conclusions in
the above scenarios depend on processes which are not well
understood, so it is still unclear whether SMBHBs could coalesce
within a Hubble time using purely stellar-dynamical processes. In
gas-rich systems it has been argued that the interaction between
binary BHs and the gaseous environment could efficiently drive the
binary to the gravitational radiation dominant domain within a
timescale of $\sim10^7~{\rm yr}$
\citep{ivanov99,gould00,armitage02,escala05,kazantzidis05}.

Although many physical processes have been proposed in the
literature to boost the hardening rates of SMBHBs, the abundance of
binary BHs in galaxy centers has not yet been strongly constrained
observationally. Because of their compactness, bound SMBHBs are very
difficult to resolve directly with current telescopes, and only
unbound SMBHBs in a couple of gas-rich (wet) galaxy mergers have
been identified \citep{komossa06}. So far all the evidence for bound
and hard SMBHBs are indirect and model dependent. The prototype
evidence for SMBHBs in AGNs is the helical morphology of radio jets
in many radio galaxies which may be due to the precession of jet
orientation in a SMBHB system \citep{begelman80}. It has been
suggested that the periodic outbursts observed in some AGNs
\citep{sillanpaa88,liu95,liu97,raiteri01} are due to the orbital
motion of the jet-emitting BH in SMBHB \citep{villata99,ostorero04},
the precession of spin axis of the rotating primary BH
\citep{romero00}, the interaction between SMBHB and a standard
accretion disk \citep{sillanpaa88,valtoja00} or an advection
dominated accretion flow (ADAF) \citep{liu02,liu06}. X-shaped
features in a subclass of radio galaxies have been attributed to the
interaction and alignment of SMBHBs and standard accretion disks
\citep[][alternatives see \citealt{merritt02,zier05}]{liu04}. There
is also some evidence for the coalescence of SMBHBs in AGNs.
\citet{liu03} suggested that the double-double radio galaxies and
the restarting jet formation in some radio galaxies are due to the
removal and refilling of inner accretion disk because of the
interaction with SMBHBs. Although there is much circumstantial
evidence for both active and coalesced SMBHBs in AGNs, it is still
unclear what fraction of SMBHBs would coalesce during or before the
AGN phase and how many could survive to the later dormant or
inactive phase. Because present observations cannot give constraints
on the physical processes which have been proposed to boost the
hardening rates of SMBHBs, \citet{liu07} suggested  measuring the
acceleration of jet precession as a way of determing the hardening
rates of SMBHBs in AGNs and to place strong constraints on models of
SMBHB evolution.  In inactive galaxies or gas-poor mergers the
observational evidence for SMBHBs is very rare.  One possible way to
identify uncoalesced SMBHBs in inactive galaxies is to detect
hypervelocity binary stars ejected by SMBHBs with the three-body
sling-shot effect \citep{lu07}. However, this method cannot be
applied to distant galaxies. Therefore, in this paper we suggest a
way to determine statistically whether SMBHBs in nearby galaxies
have coalesced.

One inevitable impact of a non-rorating SMBH on its stellar
environment is that stars passing by the BH as close as the tidal
radius
\begin{equation}
r_t \simeq r_*\left(\frac{M_{\bullet}}{M_*}\right)^{1/3} \simeq
10^{-5}~{\rm
pc}~\left(\frac{r_*}{r_\odot}\right)\left(\frac{M_*}{M_\odot}\right)^{-1/3}
\left(\frac{M_\bullet}{10^8M_\odot}\right)^{1/3}\label{rt}
\end{equation}
would be tidally disrupted \citep{hills75,rees88}, where $r_*$ and
$M_*$ are the radius and mass of the stars and $M_{\bullet}$ is the
BH mass. Part of the stellar debris will be spewed to highly
eccentric bound orbits and later fall back onto the BH, giving rise
to an outburst decaying within months to years \citep{rees88}. These
tidal flares are definitive evidences of SMBHs in inactive galaxies
\citep{komossa99,komossa04,halpern04,gezari06}. Since the frequency
of stellar disruption depends on the surrounding stellar
distribution \citep[][thereafter MT99]{syer99,magorrian99}, the
stellar disruption rate can be used as a probe of the inner
structure of galactic nucleus. For nearby early-type galaxies recent
theoretical calculation with single BH without taking into account
general relativistic (GR) effects gives an averaged stellar
disruption rate of $\sim10^{-4}~{\rm
  yr^{-1}}$ per galaxy if the stellar distribution is spherically
symmetric \citep{wang04} or possibly $1-2$ orders of magnitude
higher if the stellar distribution is non-spherical
\citep[axisymmetric or triaxial,][]{merritt04}.  However, if
SMBHBs reside in galactic centers the stellar distribution would
be dramatically changed due to the slingshot effect, and the
stellar disruption rate would be very different.  \citet{ivanov05}
studied the interaction between SMBHB and dense galactic cusp
($\sim 1$ pc) for a bound, but non-hard, binary at the dynamical
friction stage and found a high disruption rate of $10^{-2}-1~{\rm
yr^{-1}}$. \citet{merritt05} considered the stellar disruption by
single BHs immediately after SMBHB coalescence. They found that
the tidal disruption rate is significantly suppressed due to the
low density core forming during the hard stage of SMBHBs.  Since
the two stages investigated by \citet{ivanov05} and
\citet{merritt05} are two short phases in the evolution of SMBHBs
and binary BHs with mass ratios $\ga10^{-3}$ spend most of their
lifetime on the hard stage (Y02), it would be very interesting to
calculate the tidal disruption rate of stellar objects for hard
SMBHBs. Therefore, in this paper we calculate tidal disruption
rates for hard and stalling SMBHBs. We suggest that the comparison
of expected tidal disruption rate by single SMBHs or hard SMBHBs
and the future survey of X-ray or UV flares at nearby galactic
centers could tell us whether most of the SMBHs in nearby galaxies
are single or binary and whether the SMBHBs would coalesce
rapidly, as expected by some models.

The stellar disruption rate in a galaxy harboring a SMBHB depends
on: (1) the rate that stars are fed from large-scale to the
vicinity of SMBHB and (2) the probability that a star moving in
the vicinity of the SMBHB passes close enough to one of the BHs to
be tidally disrupted. Although there has been much work on the
former part, intended to study the evolution timescale of SMBHB
\citep[Q96; Y02;][]{milosavljevic03}, the latter is largely
unaddressed in the literature. Because of the chaotic nature of
three-body interaction, numerical scattering experiments are
needed to give the probability of tidal disruption. Previous
numerical experiments on star-SMBHB interactions mainly focus on
energy and angular momentum exchanges \citep[Q96;][]{sesana06} and
the limited particle numbers are insufficient to tackle the rare
events of very close encounters such as tidal disruptions.
Therefore, we begin with intensive numerical scattering
experiments (normally $\sim10^7$ particles in each run) to
calculate the tidal disruption cross sections for hard SMBHBs.
Then we apply our results to a sample of nearby galaxies and
estimate the tidal disruption rates of stellar objects both by
single and binary BHs in different types of galaxies.

This paper is organized as follows. In \S~\ref{bg} we briefly review
some basic properties of hard SMBHBs and show how to calculate
stellar disruption rate. We describe our numerical scattering
experiments in \S~\ref{experi} and present the results in
\S~\ref{results}. In \S~\ref{apply} we calculate stelar disruption
rates with realistic galaxy models and discuss the differences
between the tidal disruption rates for single and binary BHs. We
also estimate the density of tidal disruption events in local
universe and make predictions for future surveys of UV/X-ray flares
in \S~\ref{prediction}. In \S~\ref{discussion} we discuss the
observational signatures of SMBHBs in inactive galaxies and the
implications for the evolution of binary BHs. Finally we give our
conclusions in \S~\ref{conclusion}.

\section{PROPERTIES OF HARD SMBHB SYSTEMS}
\label{bg}

\subsection{Loss Cone Filling Rate}

A SMBHB with masses $M_1$ and $M_2$ ($M_1>M_2$) becomes hard when
the semimajor axis $a$ decreases to
\begin{equation}
      a\la a_h=\frac{G\mu}{4\sigma_*^2}\,,\label{ah}
\end{equation}
(Q96) where $a_h$ is the hardening radius, $\sigma_*$ is the
one-dimensional stellar velocity dispersion of background stars,
and $\mu=M_1M_2/M_{12}$ is the reduced mass with $M_{12}=M_1+M_2$.
Observations of nearby galaxies show that BH mass tightly
correlates with bulge velocity dispersion
\citep{ferrarese00,gebhardt00}. If the total BH mass $M_{12}$ of
SMBHB also follows the empirical $M_{\bullet}-\sigma_*$ relation:
$M_{12}=10^{8.13}{M_{\odot}}(\sigma_*/200~{\rm
  km~s^{-1}})^{4.02}$ \citep{tremaine02}, equation~(\ref{ah}) reduces to
\begin{equation}
    a_h\simeq3.13~{\rm pc}\frac{q}{(1+q)^2}M_8^{1/2}\,,\label{ahpc}
\end{equation}
where $M_8=M_{12}/10^8M_{\odot}$ and $q=M_2/M_1$ is the mass ratio
of the binary. In this paper we only consider $q\geqslant0.01$
because the dynamical friction timescale for SMBHBs with $q\ll0.01$
is very long (Y02).

A hard SMBHB loses energy and becomes more bound through three-body
interaction \citep{hut83}. As a result stars intruding a sphere of
radius $\sim a$ about the mass center of the binary (if not tidally
disrupted or swallowed by the BHs) will be eventually expelled with an
average energy gain
\begin{eqnarray}
    \Delta E=-\frac{GM_{12}M_*\Delta a}{2a^2}\simeq KG\mu M_*/a\,, \label{deltae}
\end{eqnarray}
where $K$ is a dimensionless factor about $1.6$ (Q96; Y02). In this
paper, unless noted otherwise, we assume that galaxies are
spherical. In a spherical potential there is just one family of
orbits (centropilic loops) and any such orbit can be labelled by its
four integrals of motion, namely the specific angular momentum $\bf
J$ and specific binding energy $\E$.  We define the \textit{loss
cone} to consist of all orbits with pericenters less than $a$, which
results in a conical region with the boundary given by $J_{\rm
lc}^2(\E,a)=2a^2[GM_{12}/a-\E]$.  In a non-spherical galaxy angular
momentum is no longer conserved and there can be more than one orbit
family.  Nevertheless, instead of a ``loss cone'' one can still
define a ``loss zone'' inside which stars will be delivered to the
SMBHB.

Mechanisms such as two-body relaxation tend to refill the loss cone
so the decay of SMBHB continues. The stars refilled into loss cone
are originally far from the SMBHB so that
\begin{equation}
    \E\ll GM_{12}/a\label{energy}
\end{equation}
and
\begin{eqnarray}
    J_{\rm lc}^2(a)\simeq2GM_{12}a\label{jlc}\,.
\end{eqnarray}
If the characteristic change of $J$ during one orbital period  of a
star is much smaller than $J_{\rm lc}$, the loss cone refilling
behaves like a diffusion process in which the loss cone filling rate
is not sensitive to $J_{\rm lc}^2$ or $a$ \citep[diffusive
regime,][]{cohn78}. Otherwise, the stars entering the loss cone have
a uniform distribution with respect to $J^2$ so that the loss cone
filling rate is proportional $J_{\rm lc}^2$ or $a$ (pinhole regime).

In inactive galaxies gas dynamics is probably not important, so a
hard binary loses energy mainly through interacting with loss cone
stars. The evolution timescale $t_h$ at this hard stage depends on
the loss cone filling rate $F^{\rm lc}(a)$, the number of stars fed
into a sphere of radius $a$ about the mass center of the binary per
unit time. From equation~(\ref{deltae}) and $d(GM_{12}/2a)/dt=F^{\rm
  lc}(a)\Delta E$ we have
\begin{eqnarray}
    t_h=|a/\dot{a}|=M_{12}/[2KF^{\rm lc}(a)m_*]\label{th}
\end{eqnarray}
(see also Y02), which implies that about an amount of $F^{\rm
lc}(a)t_h\simeq M_{12}/2K$ of stars should be consumed to
reduce $a$ by a single $e$-fold, in agreement with recent simulations
\citep{merritt06}.

\subsection{Tidal Disruption Cross Section}
\label{cstheory}

A fraction of the stars fed into the loss cone of SMBHB would be
scattered to the vicinity of a BH and get tidally disrupted. The
ratio of the tidal radius of the primary BH and the separation of
the binary is about
\begin{eqnarray}
   \frac{r_{t1}}{a_h}&\simeq&
    \nonumber
    3.71\times10^{-5}M_8^{-2/3}\left(\frac{r_*}{r_\odot}\right)\left(\frac{M_*}{M_\odot}\right)^{-1/3}
    \frac{(1+q)^{5/3}}{q_{-1}}\left(\frac{\sigma}{200~\mathrm{km~
    s^{-1}}}\right)^{2}\\
    &\simeq&3.20\times10^{-5}M_8^{-1/6}\left(\frac{r_*}{r_\odot}\right)\left(\frac{M_*}{M_\odot}\right)^{-1/3}\frac{(1+q)^{5/3}}{q_{-1}}\label{rtah}
\end{eqnarray}
(from eq.~[\ref{rt}] and [\ref{ahpc}]), which is small for typical
hard SMBHBs This implies that most stars in the loss cone would be
expelled instead of being tidally disrupted.

To further quantify the probability of tidal disruption, we
introduce $\Sigma_i(r)$ ($i=1,2$), the cross section for stars to
be scattered to a distance less than $r$ from the $i$th BH. The
probability that a loss cone star is disrupted by the $i$th BH
estimates $\Sigma_i(r_{ti})/\Sigma_i(a)$, where $r_{ti}$ is the
tidal radius of the $i$th BH. Then the stellar disruption rate is
related to the loss cone filling rate and the tidal disruption
cross section as
\begin{eqnarray}
    \dot{N}^{d}_i\simeq F^{\rm lc}(a)\frac{\Sigma_i(r_{ti})}{\Sigma_i(a)}\,.\label{nd}
\end{eqnarray}

For a single BH the geometrical cross section can be written out
analytically as
\begin{eqnarray}
    \Sigma(r)=\pi r^2\left(1+\frac{GM_{\bullet}/r}{v_0^2/2}\right)\,,\label{singlecs}
\end{eqnarray}
which increases with the gravitational potential $GM_{\bullet}/r$
and decreases with stars' kinetic energy $v_0^2/2$ at infinity. For
bound stars equation~(\ref{singlecs}) is still valid as long as the
stars fall to the BH on near-radial orbits, and in the bound case
$v_0$ is the velocity at apocenter. When gravitational focusing is
important ($GM_\bullet/r\gg v_0^2/2$) the cross section scales as
$r$. In SMBHB systems the presence of a companion BH tends to
increase the term $GM_{\bullet}/r$ by deepening the gravitational
potential and also increase the term $v_0^2/2$ by inducing orbital
motion which raises the relative velocity between each binary member
and ambient stars. Therefore in the rest frame of the primary
member, the equivalent increase in $GM_1/r$ is $GM_2/r$ and the
equivalent increase in $v_0^2/2$ is about $q^2GM_{12}/a$. So the
cross section of the primary BH is approximately
\begin{eqnarray}
      \tilde{\Sigma}_{1}(r)=\pi r^2\left(1+\frac{GM_{12}/r}{v_0^2/2+\eta
      q^2GM_{12}/a}\right)\label{appcs}
\end{eqnarray}
where $\eta$ is a correction factor of order unity. We will see
below that for different $q$s the best fit is obtained when
$\eta=1/(1+q)$. For the secondary, the roles of $M_1$ and $M_2$
exchange, so the cross section can be obtained by replacing $q$ in
equation~(\ref{appcs}) with $1/q$. Quinlan has separately explained
the behavior of the cross sections at $r/a\gg q$ and $r/a\ll q$ from
another way of understanding (Q96), while our explanation and
empirical formula are general for any $r/a$.

Because of the effect of GR equation~(\ref{appcs}) is no longer
valid when $r\sim r_{\rm S}$, where $r_{\rm S}=2GM_\bullet/c^2$ is
the Schwarzschild radius. Since the ratio of $r_{\rm S}$ and $r_t$
is
\begin{eqnarray}
  \frac{r_{\rm S}}{r_{t}} &\simeq& 0.91M_8^{2/3}\left(\frac{r_*}{r_\odot}\right)^{-1}\left(\frac{M_*}{M_\odot}\right)^{1/3}\,,\label{rsrt}
\end{eqnarray}
GR effect is important when $M_{\bullet}>3.6\times10^6M_{\odot}$
($r_t<10r_{\rm S}$ for solar type stars). In practice, one can use
the pseudo-Newtonian potential
\begin{eqnarray}
    \psi(r)=-\frac{GM_{\bullet}}{r-r_{\rm S}}\label{grp}
\end{eqnarray}
\citep{paczynski80} to simulate the GR effect so
equation~(\ref{appcs}) becomes
\begin{eqnarray}
      \tilde{\Sigma}_{1}(r)=\pi r^2\left(1+\frac{GM_{12}/(r-r_{\rm S1})}{v_0^2/2+\eta
      q^2GM_{12}/a}\right)\label{grappcs}
\end{eqnarray}
(notice that $a\gg r_{{\rm S}i}$), equivalent to multiplying the
non-relativistic cross section by a correction factor $r/(r-r_{\rm
S})$. Equation~(\ref{grappcs}) diverges when $r\rightarrow r_{\rm
S}$, but since a star plunging into the sphere of marginally bound
radius $2r_{\rm S}$ about a BH eventually falls into to the event
horizon, the cross section at $[r_{\rm S},2r_{\rm S}]$ should be
constant, equal to the cross section at $r=2r_{\rm S}$.

Both equation~(\ref{appcs}) and (\ref{grappcs}) are evaluated at
the first time that a star passes the binary. In some cases a star
could be scatted onto temporally bound orbit and encounters with
the binary many times before expelled or disrupted
(\citealt{hills83}; Q96). During such multi-encounters the
probability of tidal disruption is expected to be enhanced. Since
it is very difficult to derive analytical cross sections for these
complicated encounters, scattering experiments are needed. Stars
on near-radial orbits have $v_0^2\ll GM_{12}/a$, then
equations~(\ref{rtah}), (\ref{appcs}), and (\ref{grappcs}) suggest
that the probability of tidally disruption scales as $r_{ti}/a$,
which is of order $10^{-4}-10^{-6}$. Therefore, to get
statistically meaningful tidal disruption cross sections a large
number of particles should be used in the scattering experiments.

\section{METHOD}\label{experi}

\subsection{Scattering Experiments}

In this section we describe our numerical scattering experiments of
the restricted three-body problem. The method adopted here is
similar to those in Q96 and \citet{sesana06} but the number of test
particles in our experiments is much larger, typically $\sim10^7$,
to give statistically meaningful cross sections at
$10^{-6}<r/a<10^{-4}$.

In the scattering experiments, instead of using $\E$, the specific
binding energy of a particle in the combined potential of stars and
BHs, we use a more convenient parameter $E=GM_{12}/r-v^2/2$, the
specific binding energy about the BH binary, where $r$ is the radius
about the mass center and $v$ is the velocity of particle. We denote
the initial specific binding energy with $E_0$ and initial velocity
with $v_0$. In our problem the relevant energy range is
$-qGM_{12}/8a\la E_0\ll GM_{12}/a$ (from eq.~[\ref{ah}] and
eq.~[\ref{energy}]).

In each scattering experiment, the origin is chosen at the mass
center of the binary. In the case $-qGM_{12}/8a\la E_0<0$ (unbound
case), initially particles come form infinity
$\mathbf{r}=(x,y,z)=(b,0,\infty)$ with asymptotic velocity
$\mathbf{v}=(v_x,v_y,v_z)=(0,0,-v_0)$, where $b$ is the impact
parameter, the minimum separation between particle and mass center
if the particle feels no gravitational field. The fiducial value
for $v_0$ is $10^{-3}\sqrt{GM_{12}/a}$, to reproduce Fig. 3 in
Q96. Given $v_0$, $b^2$ is uniformly sampled in the range
$[0,b_{\rm max}^2]$, where $b_{\rm max}=\sqrt{2}J_{\rm
lc}(a)/v_0$. Particles with $b>b_{\rm max}$ hardly reach $r<a$
therefore contribute little to the tidal disruption cross section
(see \S~\ref{jtheta}). In the other case $0\leqslant E_0\ll
GM_{12}/a$ (bound case), initially we put particles at
$\mathbf{r}=(x,y,z)=(0,0,r_0)$ with an isotropic velocity
distribution ($v_0=\sqrt{2(GM_{12}/r_0-E_0)}$). $r_0$ satisfies
$a\ll r_0<GM_{12}/E_0$ because in realistic SMBHB systems most
stars enter the loss cone on near-radial orbits.

Another six parameters (four if the binary is circular) are set to
fix the initial conditions of the binary: (1) the mass ratio $q$;
(2) the eccentricity of binary orbit $e$; (3) the inclination of
orbital plane; (4) the argument of pericenter; (5) the longitude
of ascending node; (6) the initial binary phase. In each set of
experiments with fixed $q$ and $e$, the cosine of the orbital
inclination angle is evenly sampled in $[-1,1]$ and the other
three angular parameters are uniformly distributed in $[0,2\pi]$,
resulting in an isotropically filling of the loss cone.

Given the above initial conditions, the motion of a massless
particle is governed by the following coupled first-order
differential equations:
\begin{eqnarray}
  \nonumber\dot{\mathbf{r}} &=& \mathbf{v} \\
  \dot{\mathbf{v}} &=&
  -G\sum_{i=1}^{2}\frac{M_{i}(\mathbf{r}-\mathbf{r}_{i})}{|\mathbf{r}-\mathbf{r}_{i}|^3}\,,\label{accel}
\end{eqnarray}
where $\mathbf{r}_i$ is the position of the $i$th BH. In each
scattering experiment we first move a particle from its initial
position to $r_k=a(10^{8}q)^{1/4}$ along a Keplerian orbit about a
point mass $M_{12}$ at the origin. At $r_k$ the quadrupole force
from the binary is eight orders of magnitude smaller than
$GM_{12}/a^2$. Then we integrate the particle's orbit with the
subroutine \textbf{dopri8} \citep{hairer87}, an explicit
Runge-Kutta method of order (7)8. We set the threshold of
fractional error per step in $\mathbf{r}$ and $\mathbf{v}$ to be
$10^{-8}$. Raising this threshold up to $10^{-6}$ does not
significantly change our results. The integration stops if one of
the following conditions is satisfied (1) the particle leaves the
sphere of radius $r_k$ with negative binding energy; (2) the
physical integration timescale exceeds $10^{10}$ yr; (3) the
integration timestep reaches $10^6$. A small fraction ($\la0.1$
percent) of the particles are scattered to wide, bound orbits and
survive many revolutions so conditions (2) and (3) are adopted to
save computational time. We have tested our code by reproducing
Fig.~3 in Q96 and found very good agreement.

We have also used the pseudo-Newtonian potential (eq.[\ref{grp}])
in our numerical experiments to investigate the effect of GR. In
this case equation~(\ref{accel}) becomes:
\begin{eqnarray}
  \nonumber\dot{\mathbf{v}}&=&-G\sum_{i=1}^{2}\frac{M_{i}(\mathbf{r}-\mathbf{r}_{i})}{|\mathbf{r}-\mathbf{r}_{i}|(|\mathbf{r}-\mathbf{r}_{i}|-r_{{\rm
  S}i})^2}\,.
\end{eqnarray}
and we stop the integration once a particle reaches $1.01$
Schwarzschild radius about either of the BHs. In the GR experiment
the ratio of $r_{\rm S1}$ and $a_h$,
\begin{eqnarray}
  \frac{r_{\rm S1}}{a_h} &\simeq&3.0\times10^{-5}M_8^{1/2}
  \frac{1+q}{q_{-1}}\,,\label{rsah}
\end{eqnarray}
should be set in priori. For the illustrative purpose we always
set $M_8=1$. We will discuss the effect of changing $M_8$ on tidal
disruption cross sections in \S~\ref{grcs}.

\subsection{Derivation of the Cross Sections}

After each scattering experiment we record the minimum separation
between the particle and each BH. At the end of all experiments we
count the number $N_i(r)$ of particles whose minimum separations
from the $i$th BH are less than $r$. Then the normalized
multi-encounter cross section (particle are allowed to encounter
the binary as many times as they could until they are expelled) is
calculated with $\Sigma_i(r)/\Sigma_i(a)=N_i(r)/N_i(a)$. The
Poissonian error in the counts is $\sqrt{N_i(r)}$, so the error
for the normalized cross section is
$N_i(r)/N_i(a)\sqrt{1/N_i(r)+1/N_i(a)}$ and the corresponding
fractional error in statistics is $\sigma_{\rm
stat}(r)=\sqrt{1/N_i(r)+1/N_i(a)}$.

Following Q96, we also record the minimum separation between a
particle and each BH during their first encounter and derive the
single-encounter cross sections. In \S~\ref{nongrcs} we compare
the single-encounter cross sections and those from the analytical
approximations (eq.~[\ref{appcs}] and [\ref{grappcs}]).

\section{RESULTS FROM THE SCATTERING EXPERIMENTS}
\label{results}

\subsection{Non-relativistic Cross Sections}
\label{nongrcs}

\begin{figure}
  \plotone{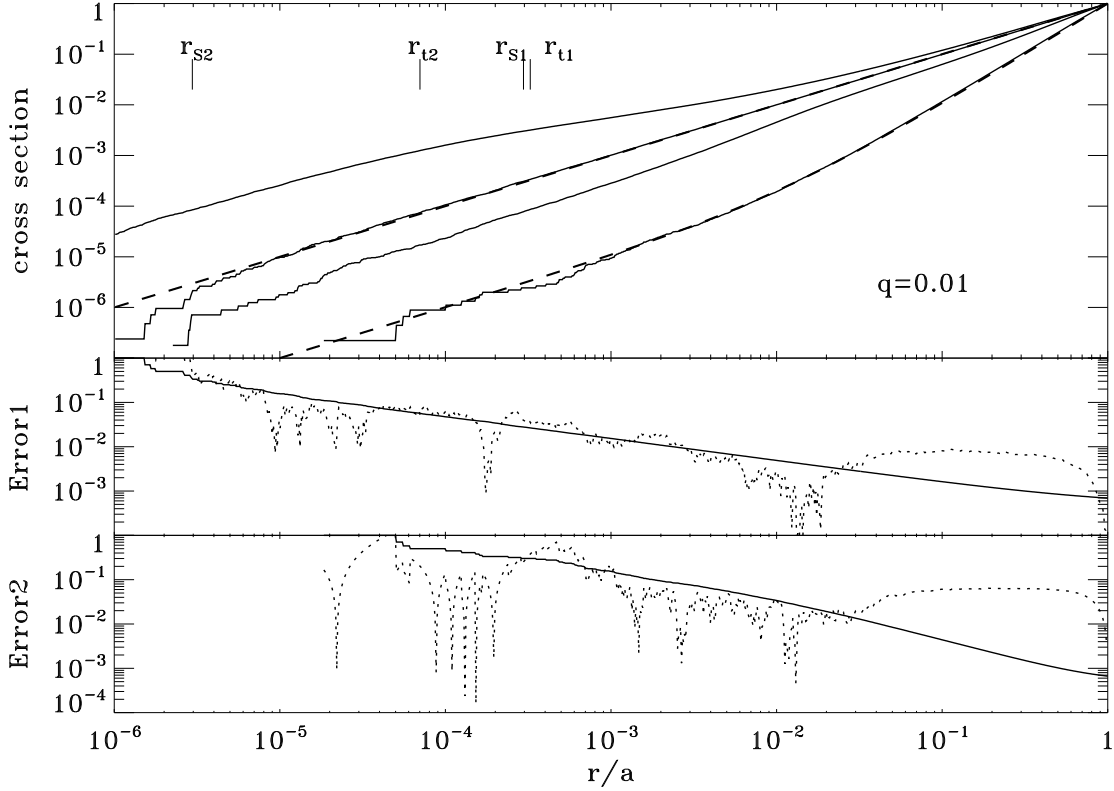}\\
\caption{\textit{Top}: Normalized multi- and single-encounter cross
sections from $10^7$ particles for $q=0.01$. From top to bottom, the
first two solid lines, respectively, refer to the multi- and
single-encounter cross sections of the primary BH, and the third and
fourth ones are of the secondary. Empirical cross sections (dashed
lines) and positions of tidal radii and Schwarzschild radii for
$M_{12}=10^8M_{\odot}$ (short vertical lines ) are also plotted.
\textit{Middle}: Fractional errors of statistical fluctuation
(solid) and empirical cross sections (dotted), for the primary BH.
\textit{Bottom}: The same as the middle panel but for the secondary
BH.}\label{cs001}
\end{figure}

First we consider the unbound case. We set
$v_0=10^{-3}\sqrt{GM_{12}/a}$ (after Fig.~3 in Q96), $q=0.01$, and
use the Newtonian potential. The normalized mullti- and
single-encounter cross sections from $N\simeq10^7$ particles are
presented as solid lines in the top panel of Figure~\ref{cs001}. The
cross sections are plotted as a function of $r/a$ so that they can
be easily scaled. The perturbations at the lower ends are due to
statistical fluctuation. In the top panel of Figure~\ref{cs001} we
also plot the empirical single-encounter cross sections (see
eq.~[\ref{appcs}]) as dashed lines. We estimate the fractional error
of the empirical formula with $\sigma_{\rm
app}\equiv|\Sigma_i(r)/\Sigma_i(a)-\tilde{\Sigma}_i(r)/\tilde{\Sigma}_i(a)|/[\Sigma_i(r)/\Sigma_i(a)]$,
where $\Sigma_i(r)/\Sigma_i(a)$ is from scattering experiments and
$\tilde{\Sigma}_i(r)$ from equation~(\ref{appcs}). $\sigma_{\rm
app}$ and $\sigma_{\rm stat}$ for the primary and secondary BHs are
shown in the middle and bottom panels of Figure~\ref{cs001}. For
both BHs we found that at $r\ll a$ $\sigma_{\rm app}$ is almost
always below $\sigma_{\rm stat}$, indicating that the empirical
cross sections agrees very well with the numerical results. At
$r>qa$, although $\sigma_{\rm app}$ is larger than $\sigma_{\rm
stat}$, the difference does not exceed $10\%$.

\begin{figure}
  \plotone{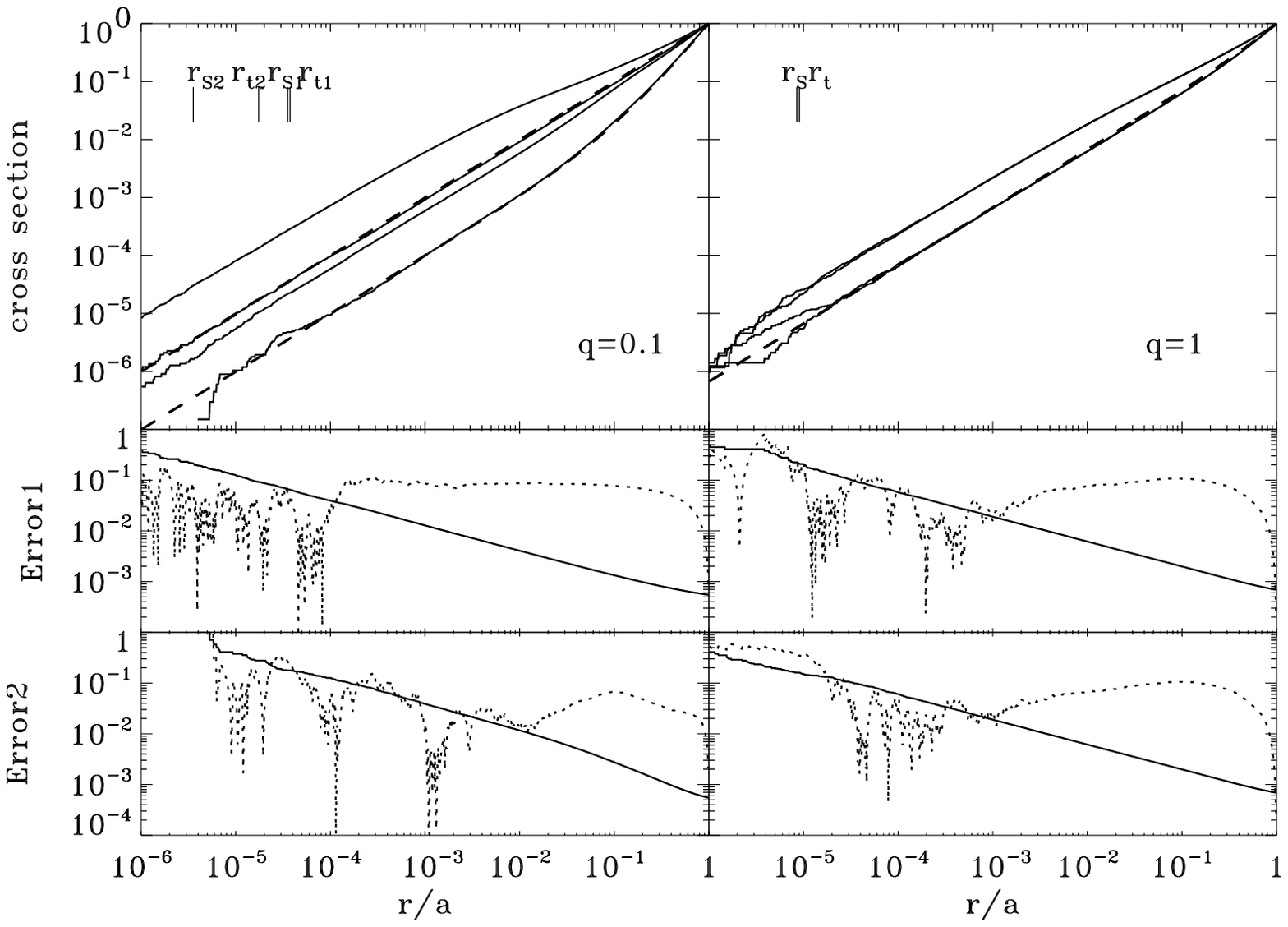}\\
\caption{Close-encounter cross sections and fractional errors for
$q=0.1$ (the left panels, $10^7$ particles) and $q=1$ (the right
panels, $10^6$ particles). Lines have the same meanings as those
in Figure~\ref{cs001}.}\label{fitnongr}
\end{figure}

For $q=0.1$ and $q=1$, the numerical (solid lines) and empirical
approximations (dashed lines) cross sections are presented in the
top left and top right panels of Figure~\ref{fitnongr}. $\sigma_{\rm
stat}$ (solid lines) and $\sigma_{\rm app}$ (dotted lines) for the
single-encounters are presented in the middle and bottom panels.
Although the approximations become worse when $q$ is large, the
difference between approximation error and statistic fluctuation is
always below $10\%$.

Figure~\ref{cs001} and Figure~\ref{fitnongr} show that at $r\la
q^2a$ both the multi- and single-encounter cross sections scales
linearly as $r/a$, implying that gravitational focusing is
important for very close encounters. The cross sections at
$r/a<10^{-6}$ could be obtained by using this scaling. It is also
clear that the normalized multi-encounter cross sections are
greater than the single-encounter ones as is predicted in
\S~\ref{cstheory}. However, we find that in our experiments
although a particle could encounter with the binary many times, no
particle more than once enters the sphere of tidal radius (for
$M_8=1$) about either BH. This is because the probability for a
particle to enter the sphere of tidal radius about the $i$th BH
$n$ times roughly scales as $[\Sigma_i(r_{ti})/\Sigma_i(a)]^n$ or
$(r_{ti}/a)^n$, which becomes extremely low for $r_{ti}\ll a$ and
$n>1$. For the similar reason, the event that a particle
successively approaches the tidal radii of the primary and
secondary BHs is also extremely rare.

\subsection{Dependence on Particle's Binding Energy}
\label{boundstars}

To investigate the effect of binding energy, we carried out two
test experiments for $E_0=-0.01$ and $E_0=0.01$ (in unit of
$GM_{12}/a$), each with $\sim10^6$ particles and $q=0.01$. In the
unbound case the asymptotic velocity of the intruding particles is
$v_0\simeq0.14\sqrt{GM_{12}/a}$, slightly greater than the limit
$0.85\sqrt{GM_{12}q/a}$ (from the fitting formula eq. 17 in Q96)
above which the binary becomes soft. In the bound case, initially
the particles are at $r_0=50a$ with an isotropic velocity of
$v_0\simeq0.14\sqrt{GM_{12}/a}$. We do not consider $E_0=0.1$
because such initial condition results in unrealistically bound
particles with $r_0\leqslant10a$.

\begin{figure}
  \plotone{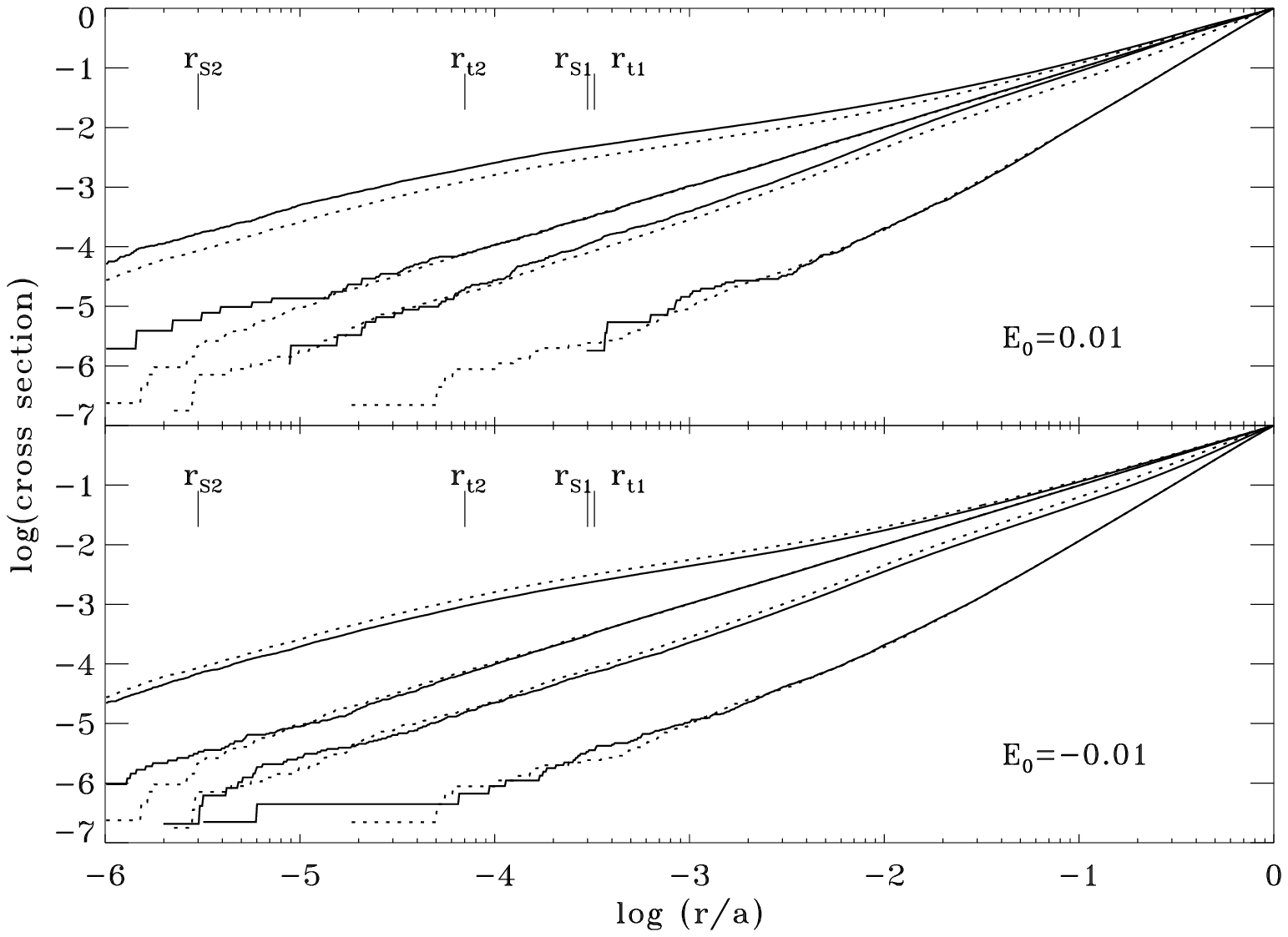}\\
\caption{Normalized multi- and single-encounter cross sections
from $\sim10^6$ particles (solid lines, $q=0.01$) for $E_0=0.01$
(\textit{top}) and $E_0=-0.01$ (\textit{bottom}). $E_0$ is in unit
of $GM_{12}/a$. Dotted lines are from the fiducial experiments
(\S~\ref{nongrcs}) and short vertical lines have the same meanings
as those in previous figures.}\label{bound}
\end{figure}

Results from these test experiments are presented in
Figure~\ref{bound} as solid lines. Cross sections for the fiducial
value $E_0\sim-10^{-6}GM_{12}/a$ (from \S~\ref{nongrcs}) are also
plotted for reference. Although the physical cross sections increase
prominently with $E_0$ which is expected by equation~\ref{appcs},
Figure~\ref{bound} shows that in both bound and unbound cases the
normalized cross sections for both multi- and single-encounters seem
not varying significantly with $E_0$. This is because when $|E_0|\ll
GM_{12}/a$, the velocity of the particles passing $r\la a$ from the
binary is not sensitive to $E_0$ but to the depth of the
gravitational potential at $r$. Therefore, different particles
passing by the same BH binary feel the similar strength of
gravitational focusing and have similar interaction cross sections
with the BH components of binary. Since the normalized cross
sections do not vary significantly with $E_0$ as long as $|E_0|\ll
GM_{12}/a$, it is reasonable to apply the fiducial cross sections to
various binaries with a wide range of semimajor axis. For example,
according to equation~(\ref{ah}), varying $E_0$ from
$-10^{-6}GM_{12}/a$ to $-10^{-2}GM_{12}/a$ corresponds to increasing
$a$ from $\sim10^{-5}q_{-1}^{-1}a_{ h}$ to $\sim 0.1q_{-1}^{-1}a_{
h}$.

\subsection{Dependence on Particle's Initial Angular Momentum}\label{jtheta}

In both experiments for $E_0\geqslant0$ and $E_0<0$ initially the
particle follow a uniform distribution in $J^2$ and the loss cone
filling rate is isotropic, i.e., the cross sections obtained in
\S~\ref{nongrcs} and \S~\ref{boundstars} have been equally
averaged over the whole loss cone and over all directions. Such
averaging may be oversimplified in realistic SMBHB systems.

\begin{figure}
  \plotone{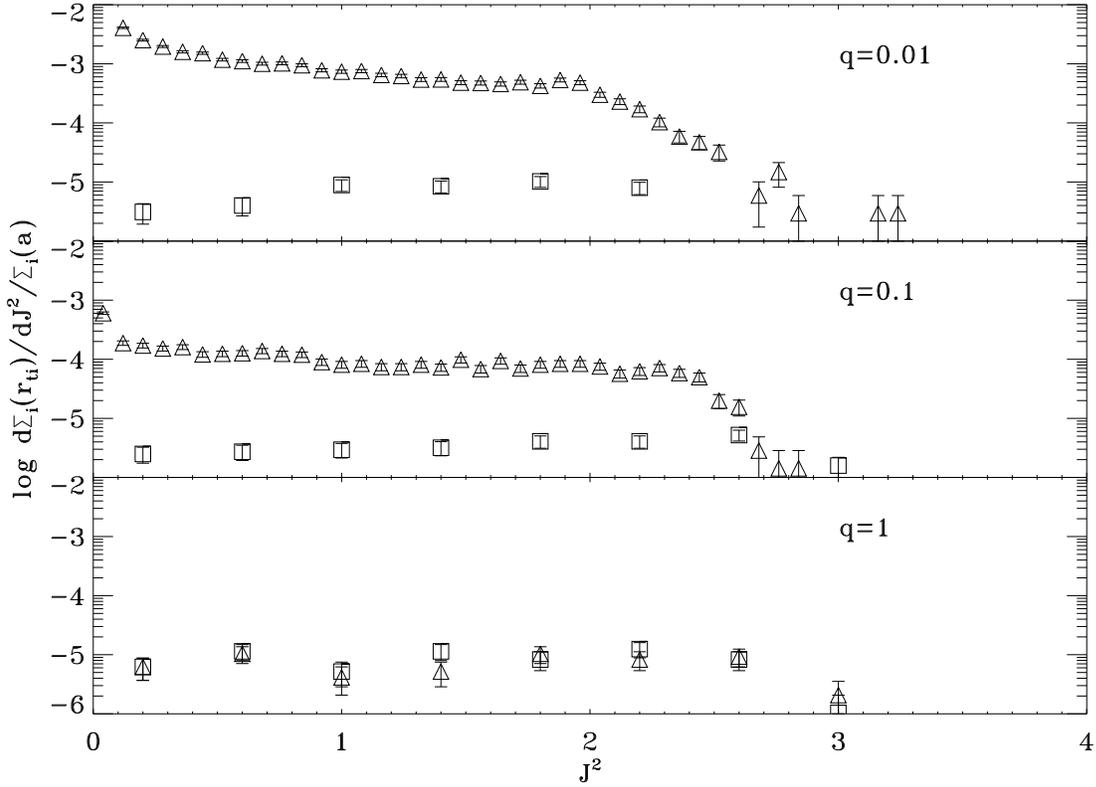}\\
\caption{Differential tidal disruption cross sections with respect
to the square of initial angular momentum (in unit of $GM_{12}a$)
for $q=0.01$ (\textit{top}), $q=0.1$ (\textit{middle}), and $q=1$
(\textit{bottom}). Triangles and squares, respectively, refer to the
differential cross sections of primary and secondary BHs. Error bars
indicate the Poissonian errors in the counts of tidal disruption
events. When differential cross section is low, large perturbation
occurs due to the noise in the counts.}\label{dfjdr}
\end{figure}

To investigate the dependence of the cross sections on the
amplitude of initial angular momentum, in Figure~\ref{dfjdr} we
plot the differential tidal disruption cross sections
$d\Sigma_i(r_{ti})/dJ^2/\Sigma_i(a)$ with respect to $J^2$. The
triangles and squares, respectively, refer to the primary and
secondary BHs. $d\Sigma_i(r_{ti})/dJ^2/\Sigma_i(a)$ is obtained as
follows. For each bin of $\Delta J^2$ we count the number $\Delta
N_i(r_{ti})$ of particles whose initial angular momenta fall in
this bin and minimum separations from the $i$th BH are less than
$r_{ti}$. Here $r_{ti}$ is from equation~(\ref{rtah}) with $M_8=1$
and we only use the data for multi-encounters. Then $d\Sigma/dJ^2$
is calculated with $\Delta N_i(r_{ti})/\Delta J^2/N_i(a)$ so that
the integration over $J^2$ results in the normalized cross section
of tidal disruption. The Poissonian errors in the counts have been
indicated by the error bars in Figure~\ref{dfjdr}. Note that the
differential cross section [or $\Delta N_i(r_{ti})$ alone] derived
in this way is proportional to the probability of tidal disrupting
a particle with initial angular momentum $J$.  Generally speaking,
both differential cross sections for the primary and secondary BHs
are flat at $J^2<J_{\rm lc}^2\simeq2GM_{12}a$. The differential
cross sections start to cut off above $J_{\rm lc}^2$ and particles
with initial angular momentum greater than $2J_{\rm lc}^2$
contribute little to the cross sections. When $q\ll1$ the
differential cross section for primary BH rises steeply with
decreasing $J^2$ at $J\ll J_{\rm lc}$ but even in this case the
contribution to the total cross section from particles with larger
$J$ is still significant.

\begin{figure}
\plotone{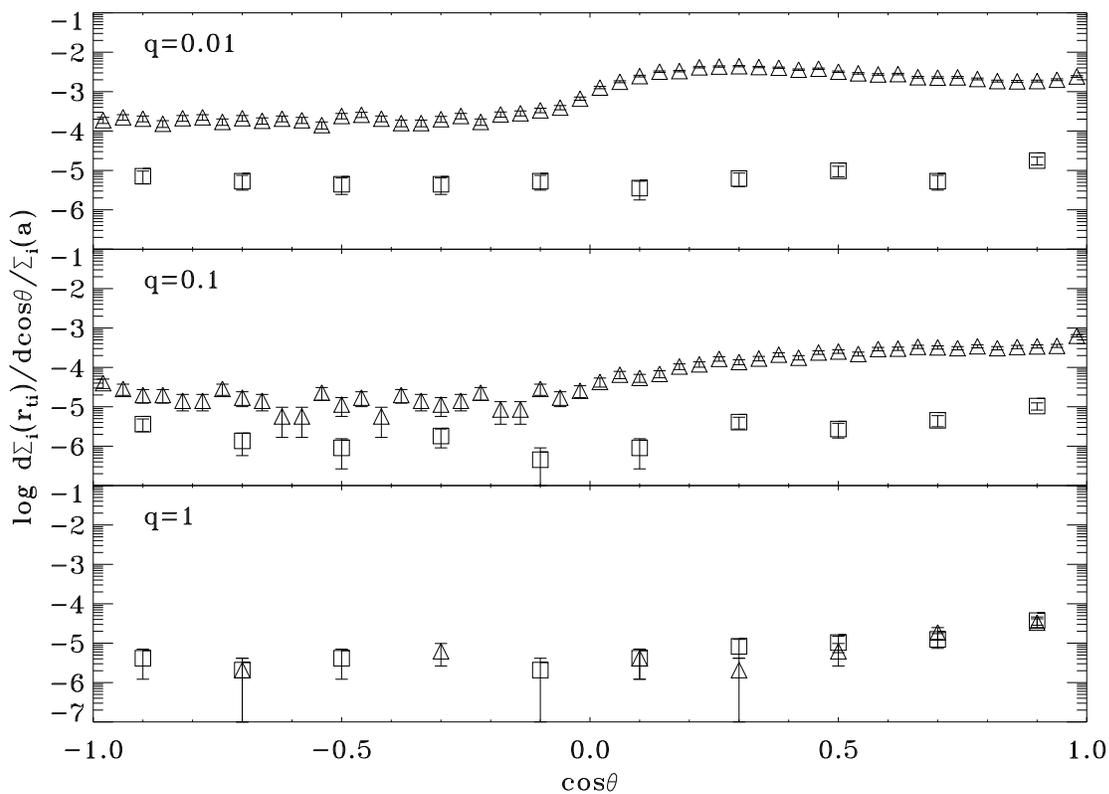}\\
\caption{Differential tidal disruption cross sections with respect
to $\cos\theta$ for $q=0.01$ (\textit{top}), $q=0.1$
(\textit{middle}), and $q=1$ (\textit{bottom}). Triangles and
squares, respectively, refer to the primary and secondary
BHs.}\label{dfthetadr}
\end{figure}

The cross sections also depend on the direction of the initial
angular momentum vector $\mathbf{J}$. We define $\theta$ as the
relative angle between $\mathbf{J}$ and binary's orbital angular
momentum. Figure~\ref{dfthetadr} shows the differential tidal
disruption cross sections with respect to $\cos\theta$. The
differential cross sections are obtained following the same
described above. When $q\ll1$ the particles on corotating orbits
($\cos\theta>0$) are more likely to be disrupted, while for
secondary BHs the dependence of the cross sections on $\theta$ is
weak. Unless the stellar distribution is extremely flattened or
the star cluster around SMBHB is significantly rotating the total
cross section would not differ much from those presented in
Figures~\ref{cs001} and \ref{fitnongr}.

\subsection{Effect of Eccentric Binary Orbit}

\begin{figure}
\plotone{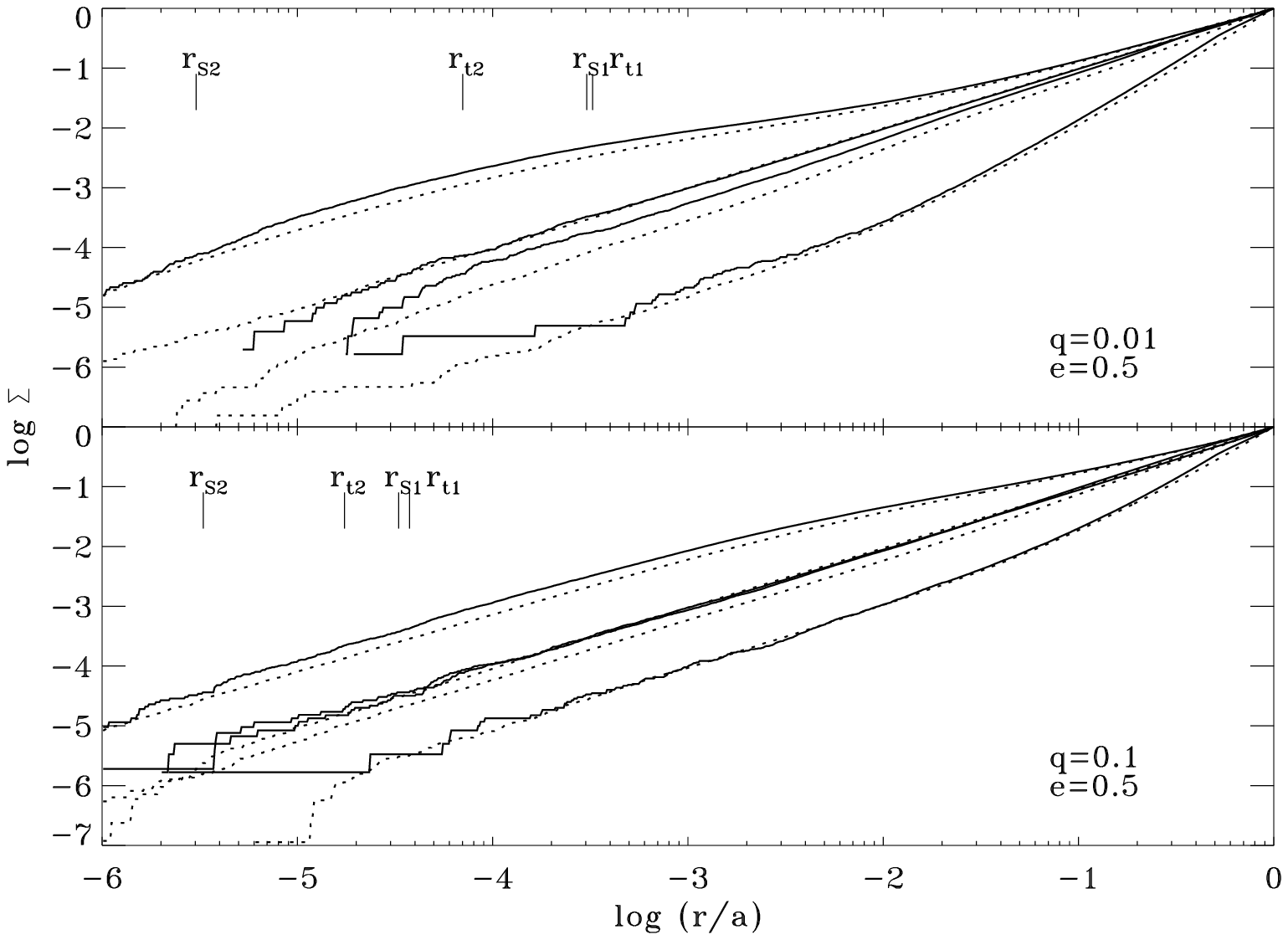} \caption{Normalized multi- and single-encounter
cross sections from $10^6$ particles (solid lines) for elliptical
binaries with $e=0.5$, $q=0.01$ (\textit{top}) or $q=0.1$
(\textit{bottom}). Dotted lines are results from the circular binary
case. Short vertical lines have the same meanings as those in
Fig.~\ref{cs001}.} \label{quilane}
\end{figure}

To see clearly the effect of the eccentricity of binary's orbit,
we set an extreme eccentricity $e=0.5$ and carried out two test
experiments for $q=0.01$ and $0.1$, each with $10^6$ particles.
The results (solid lines) are presented in Figure~\ref{quilane}.
Cross sections for $e=0$ (dotted lines, from \S~\ref{nongrcs}) are
also plotted for reference. Increasing eccentricity only slightly
increase the normalized multi-encounter cross sections for both BH
components.

\subsection{Effect of GR}
\label{grcs}

\begin{figure}
\plotone{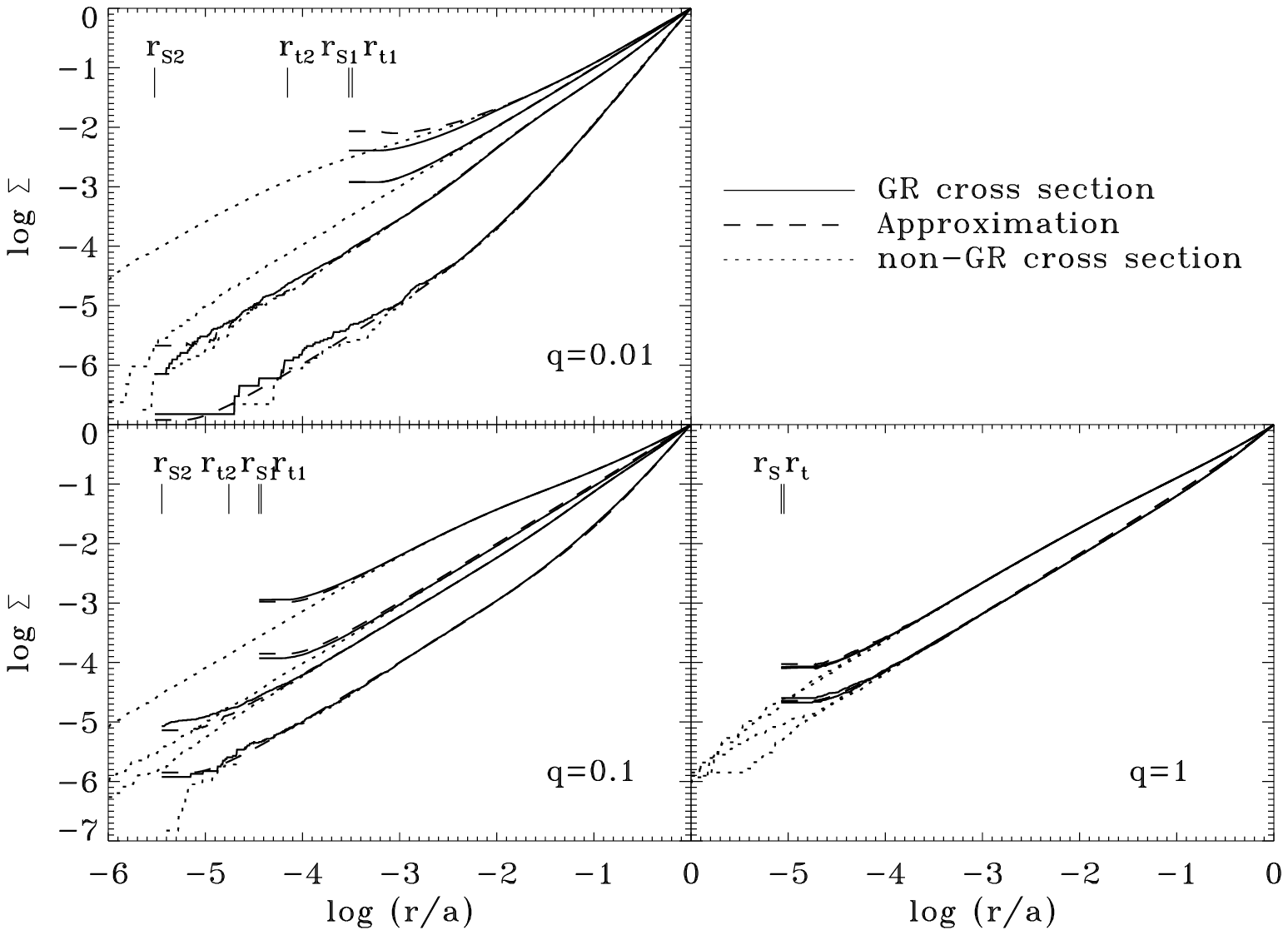}\\
\caption{Relativistic cross sections from $10^7$ particles (solid
lines) for $q=0.01$ (\textit{top left}), $q=0.1$ (\textit{bottom
left}), and $q=1$ (\textit{bottom right}). The empirical cross
sections in GR case (dashed lines) and the cross sections from
non-GR experiments (dotted lines) are also plotted. Short vertical
lines have the same meanings as those in
Fig.~\ref{cs001}.}\label{fit}
\end{figure}

To study the effect of GR we set the gravitational potential to be
pseudo-Newtonian with equation~[\ref{grp}] and $M_8=1$ and then
repeated the scattering experiments. For each $q$ the result from
$10^7$ particles is shown in Figure~\ref{fit} as solid lines.
Dotted lines are results from the non-relativistic experiments.
Below the marginally bound radius $2r_{\rm S}$ both the cross
sections for multi- and single-encounters become constant,
representing the relativistic effect.

The dashed lines are empirical cross section in the GR case. For the
single-encounters we have used equation~(\ref{grappcs}) and set the
cross sections constant between $r=r_{\rm S}$ and $2r_{\rm S}$. The
resulting cross sections agree well with the numerical ones. The
approximate cross sections to the multi-encounters are obtained by
multiplying the non-relativistic cross sections by the correction
factor $r/(r-r_{\rm S})$ and keeping the cross sections constant
below $2r_{\rm S}$. These resulting cross sections well agree with
the numerical ones as long as $r_{\rm S}\ll q^2a$, i.e., in the
gravitational focusing dominated regime. Due to GR effect the cross
sections at $r=r_{\rm S}$ are four times greater than the
non-relativistic ones.

For other values of $M_8$ the location of Schwarzschild radius
moves along the $r/a$ axis. Instead of repeating the whole
simulation we use non-GR cross sections to calculate the
approximate relativistic cross sections, following the procedure
described above.

\section{STELLAR DISRUPTION RATES IN NEARBY GALAXIES}\label{apply}

\subsection{Galaxy Sample}\label{sample}

In this section we estimate the tidal disruption rates
(eq.~\ref{nd}) of stellar objects for a sample of nearby galaxies.
The largest uncertainty in our calculation comes from the loss cone
refilling rate $F^{\rm lc}(a)$; for simplicity, and to ease
comparison with earlier work, we assume that two-body relaxation is
the only process contributing to $F^{\rm lc}(a)$.

Our galaxy sample consists of $51$ nearby elliptical galaxies which
are listed in Table~\ref{table1}. For each galaxy the BH mass is
obtained with \citeauthor{haring04}'s $M_\bullet-M_{\rm bulge}$
relation, where the bulge mass $M_{\rm bulge}$ is from the scaling
relation between stellar mass and luminosity \citep{magorrian98}.
The assumed mass to light ratios and the resulting BH masses are
given by columns 2 and 3 in Table~\ref{table1}.

\begin{deluxetable}{lcccrrrrrrrrr}
    \tablewidth{0pt}
    \tabletypesize{\scriptsize}
    \rotate
    \tablecaption{Galaxy sample\label{table1}}
    \tablehead{
    \colhead{} &
    \colhead{$\Upsilon_V$} &
    \colhead{$\log M_\bullet$} &
    \colhead{$\log F^{\rm lc}_{\rm single}$} &
    \colhead{$\log a_{\rm stall}$} &
    \colhead{$\log t_{\rm evol}$} &
    \colhead{$\log F^{\rm lc}_{\rm binary}$} &
    \colhead{$\log \dot{N}^d_1$} &
    \colhead{$\log \dot{N}^d_2$} &
    \colhead{$\log \dot{N}^d_1$} &
    \colhead{$\log \dot{N}^d_2$} &
    \colhead{$\log \dot{N}^d_1$} &
    \colhead{$\log \dot{N}^d_2$} \\
    \colhead{Name} &
    \colhead{($M_\odot/L_\odot$)} &
    \colhead{($M_\odot$)} &
    \colhead{(${\rm yr^{-1}}$)} &
    \colhead{(pc)} &
    \colhead{(yr)} &
    \colhead{(${\rm yr^{-1}}$)} &
    \colhead{(${\rm yr^{-1}}$)} &
    \colhead{(${\rm yr^{-1}}$)} &
    \colhead{(${\rm yr^{-1}}$)} &
    \colhead{(${\rm yr^{-1}}$)} &
    \colhead{(${\rm yr^{-1}}$)} &
    \colhead{(${\rm yr^{-1}}$)} \\
    \colhead{(1)} &
    \colhead{(2)} &
    \colhead{(3)} &
    \colhead{(4)} &
    \colhead{(5)} &
    \colhead{(6)} &
    \colhead{(7)} &
    \colhead{(8)} &
    \colhead{(9)} &
    \colhead{(10)} &
    \colhead{(11)} &
    \colhead{(12)} &
    \colhead{(13)}
    }
    \startdata

A2052       &   7.41    &   9.17    &   -5.89   &   1.76    &   13.55   &   -4.88   &   -9.75   &   -12.54  &   -10.33  &   -11.87  &   -11.11  &   -11.41  \\
NGC 1023    &   4.88    &   7.84    &   -4.13   &   0.25    &   9.98    &   -2.65   &   -6.47   &   -9.71   &   -7.03   &   -8.54   &   -7.59   &   -7.61   \\
NGC 1172    &   5.39    &   8.16    &   -4.21   &   1.09    &   10.50   &   -2.85   &   -7.38   &   -10.20  &   -7.97   &   -9.49   &   -8.76   &   -9.06   \\
NGC 1316    &   7.60    &   9.25    &   -4.64   &   0.89    &   12.46   &   -3.71   &   -7.72   &   -10.47  &   -8.25   &   -9.80   &   -9.04   &   -8.94   \\
NGC 1399    &   6.33    &   8.67    &   -5.28   &   0.87    &   12.73   &   -4.57   &   -8.73   &   -11.60  &   -9.27   &   -10.83  &   -10.10  &   -10.10  \\
NGC 1400    &   5.69    &   8.33    &   -4.79   &   0.97    &   11.49   &   -3.67   &   -8.05   &   -11.33  &   -8.59   &   -10.17  &   -9.57   &   -9.57   \\
NGC 1426    &   5.04    &   7.95    &   -4.44   &   0.95    &   10.38   &   -2.94   &   -7.45   &   -10.29  &   -7.97   &   -9.51   &   -8.78   &   -8.78   \\
NGC 1600    &   7.44    &   9.19    &   -5.80   &   1.33    &   13.81   &   -5.13   &   -9.57   &   -12.49  &   -10.13  &   -11.70  &   -10.97  &   -10.91  \\
NGC 1700    &   6.25    &   8.63    &   -4.37   &   1.18    &   10.84   &   -2.71   &   -7.22   &   -10.06  &   -7.74   &   -9.24   &   -8.55   &   -8.55   \\
NGC 221     &   2.71    &   5.96    &   -3.68   &   -3.34   &   8.52    &   -3.16   &   -4.96   &   -6.48   &   -4.84   &   -6.07   &   -5.37   &   -5.37   \\
NGC 224     &   4.62    &   7.67    &   -4.54   &   -0.47   &   10.91   &   -3.75   &   -7.00   &   -9.81   &   -7.49   &   -8.98   &   -8.04   &   -8.10   \\
NGC 2636    &   3.93    &   7.15    &   -5.05   &   0.88    &   9.95    &   -3.30   &   -7.99   &   -11.30  &   -8.53   &   -10.10  &   -9.49   &   -9.49   \\
NGC 2832    &   7.76    &   9.32    &   -5.53   &   1.57    &   13.57   &   -4.75   &   -9.41   &   -12.69  &   -9.94   &   -11.44  &   -10.95  &   -10.95  \\
NGC 2841    &   4.66    &   7.69    &   -4.59   &   0.74    &   10.21   &   -3.02   &   -7.36   &   -10.68  &   -7.92   &   -9.46   &   -8.92   &   -8.92   \\
NGC 3115    &   5.39    &   8.16    &   -4.03   &   0.56    &   10.19   &   -2.53   &   -6.56   &   -9.59   &   -7.12   &   -8.62   &   -7.87   &   -7.80   \\
NGC 3377    &   4.53    &   7.60    &   -3.85   &   -0.40   &   10.04   &   -2.94   &   -6.28   &   -9.07   &   -6.77   &   -8.26   &   -7.37   &   -7.36   \\
NGC 3379    &   5.21    &   8.05    &   -4.93   &   0.63    &   11.48   &   -3.93   &   -8.07   &   -10.89  &   -8.62   &   -10.13  &   -9.31   &   -9.31   \\
NGC 3599    &   4.55    &   7.61    &   -4.61   &   0.92    &   10.16   &   -3.05   &   -7.63   &   -10.41  &   -8.18   &   -9.70   &   -9.04   &   -9.26   \\
NGC 3605    &   4.13    &   7.31    &   -4.78   &   0.44    &   10.00   &   -3.19   &   -7.36   &   -10.26  &   -7.94   &   -9.42   &   -8.79   &   -8.87   \\
NGC 3608    &   5.48    &   8.21    &   -4.79   &   0.94    &   11.12   &   -3.41   &   -7.80   &   -11.07  &   -8.35   &   -9.91   &   -9.31   &   -9.31   \\
NGC 4168    &   6.39    &   8.70    &   -5.56   &   1.20    &   12.77   &   -4.58   &   -9.09   &   -11.93  &   -9.61   &   -11.11  &   -10.45  &   -10.39  \\
NGC 4239    &   3.50    &   6.78    &   -5.24   &   -0.02   &   10.02   &   -3.74   &   -7.67   &   -10.47  &   -8.19   &   -9.70   &   -8.90   &   -8.90   \\
NGC 4365    &   6.71    &   8.86    &   -5.24   &   0.96    &   12.82   &   -4.46   &   -8.65   &   -11.50  &   -9.21   &   -10.69  &   -10.06  &   -10.14  \\
NGC 4387    &   3.95    &   7.17    &   -4.83   &   0.06    &   10.02   &   -3.36   &   -7.26   &   -10.09  &   -7.78   &   -9.25   &   -8.40   &   -8.49   \\
NGC 4434    &   4.01    &   7.22    &   -4.48   &   -0.37   &   9.98    &   -3.27   &   -6.74   &   -9.42   &   -7.25   &   -8.74   &   -7.86   &   -7.85   \\
NGC 4458    &   4.01    &   7.22    &   -4.31   &   -0.39   &   9.99    &   -3.28   &   -6.73   &   -9.43   &   -7.25   &   -8.73   &   -7.86   &   -7.84   \\
NGC 4464    &   3.54    &   6.82    &   -3.99   &   -2.38   &   9.78    &   -3.56   &   -5.73   &   -7.62   &   -5.79   &   -7.13   &   -6.40   &   -6.40   \\
NGC 4467    &   2.92    &   6.20    &   -4.52   &   -2.89   &   9.59    &   -3.99   &   -6.01   &   -7.73   &   -5.96   &   -7.27   &   -6.55   &   -6.55   \\
NGC 4472    &   7.29    &   9.12    &   -5.30   &   0.83    &   13.29   &   -4.67   &   -8.66   &   -11.43  &   -9.20   &   -10.72  &   -9.95   &   -9.89   \\
NGC 4478    &   4.49    &   7.58    &   -4.78   &   0.80    &   10.27   &   -3.20   &   -7.66   &   -10.55  &   -8.20   &   -9.76   &   -9.03   &   -8.97   \\
NGC 4486    &   7.05    &   9.02    &   -5.57   &   0.83    &   13.40   &   -4.89   &   -7.27   &   -10.08  &   -7.76   &   -9.25   &   -8.34   &   -8.40   \\
NGC 4486b   &   3.18    &   6.48    &   -5.11   &   -0.89   &   10.00   &   -4.03   &   -8.91   &   -11.65  &   -9.45   &   -10.98  &   -10.24  &   -10.14  \\
NGC 4551    &   4.10    &   7.28    &   -4.70   &   0.01    &   10.00   &   -3.23   &   -7.03   &   -10.27  &   -7.56   &   -9.09   &   -8.18   &   -8.19   \\
NGC 4552    &   5.66    &   8.32    &   -4.81   &   0.82    &   11.79   &   -3.98   &   -8.19   &   -11.03  &   -8.77   &   -10.32  &   -9.58   &   -9.88   \\
NGC 4564    &   4.72    &   7.73    &   -4.24   &   0.23    &   10.00   &   -2.77   &   -6.61   &   -9.86   &   -7.16   &   -8.66   &   -7.79   &   -7.88   \\
NGC 4570    &   4.80    &   7.79    &   -4.11   &   -0.20   &   9.97    &   -2.69   &   -6.16   &   -8.93   &   -6.67   &   -8.14   &   -7.25   &   -7.23   \\
NGC 4621    &   5.88    &   8.43    &   -4.11   &   0.82    &   10.53   &   -2.60   &   -6.78   &   -9.65   &   -7.35   &   -8.83   &   -8.14   &   -8.14   \\
NGC 4636    &   6.28    &   8.65    &   -5.52   &   0.80    &   12.95   &   -4.81   &   -8.93   &   -11.85  &   -9.49   &   -11.00  &   -10.14  &   -10.14  \\
NGC 4649    &   6.80    &   8.90    &   -5.37   &   0.82    &   13.09   &   -4.70   &   -8.75   &   -11.76  &   -9.30   &   -10.79  &   -10.04  &   -9.97   \\
NGC 4697    &   5.64    &   8.30    &   -4.69   &   0.66    &   11.05   &   -3.25   &   -7.34   &   -10.61  &   -7.90   &   -9.45   &   -8.59   &   -8.55   \\
NGC 4742    &   4.18    &   7.35    &   -3.50   &   -1.98   &   9.80    &   -3.05   &   -5.35   &   -7.34   &   -5.48   &   -6.85   &   -6.11   &   -6.10   \\
NGC 4874    &   8.57    &   9.64    &   -6.07   &   1.61    &   14.50   &   -5.37   &   -9.92   &   -12.73  &   -10.51  &   -12.06  &   -11.29  &   -11.59  \\
NGC 4889    &   8.33    &   9.54    &   -5.77   &   1.60    &   14.13   &   -5.09   &   -9.67   &   -12.45  &   -10.24  &   -11.78  &   -11.07  &   -11.30  \\
NGC 524     &   6.13    &   8.57    &   -5.10   &   1.00    &   12.13   &   -4.06   &   -8.38   &   -11.73  &   -8.94   &   -10.45  &   -9.84   &   -9.97   \\
NGC 5813    &   6.44    &   8.73    &   -5.14   &   1.10    &   12.51   &   -4.29   &   -8.67   &   -11.95  &   -9.21   &   -10.79  &   -10.09  &   -10.21  \\
NGC 5845    &   4.66    &   7.69    &   -4.41   &   0.78    &   10.00   &   -2.81   &   -7.20   &   -10.48  &   -7.77   &   -9.31   &   -8.73   &   -8.73   \\
NGC 596     &   5.52    &   8.24    &   -4.61   &   0.95    &   10.80   &   -3.07   &   -7.46   &   -10.73  &   -8.02   &   -9.57   &   -8.97   &   -8.97   \\
NGC 6166    &   8.46    &   9.60    &   -6.24   &   1.69    &   14.68   &   -5.59   &   -10.26  &   -13.55  &   -10.81  &   -12.39  &   -11.79  &   -11.79  \\
NGC 720     &   6.23    &   8.62    &   -5.56   &   0.96    &   13.00   &   -4.89   &   -9.17   &   -11.93  &   -9.72   &   -11.23  &   -10.57  &   -10.79  \\
NGC 7332    &   4.70    &   7.72    &   -3.73   &   -0.56   &   10.01   &   -2.80   &   -5.96   &   -8.73   &   -6.44   &   -7.90   &   -6.98   &   -7.02   \\
NGC 7768    &   7.73    &   9.31    &   -5.42   &   1.62    &   13.36   &   -4.56   &   -9.27   &   -12.19  &   -9.82   &   -11.36  &   -10.76  &   -10.76  \\

    \enddata

\tablecomments{Column 1 is the galaxy name. Column 2 is the $V$-band
mass-to-light ratio and Column 3 gives the corresponding BH mass
according to the $M_\bullet-M_{\rm bulge}$ relation
\citep{haring04}. Columns 4 is the loss cone filling rates for
single BHs in non-GR case. Columns 5, 6, and 7 list the stalling
radius, evolution timescale, and loss cone filling rate for SMBHBs
with mass ratio $q=0.1$, the results for $q=0.01$ and $1$ being
essentially identical. Both $t_{\rm evol}$ and $F^{\rm lc}_{\rm
binary}$ are evaluated at $a_{\rm stall}$, given by the minimum of
equation~\ref{tevol}.
 The tidal disruption
rates for $q=0.01$ are given in Columns 8 and 9, respectively for
the primary and secondary BHs. Columns 10 and 11 have the same
meanings as those in Columns 9 and 10, but for $q=0.1$, while
Columns 12 and 13 are for $q=1$.}

\end{deluxetable}

\subsection{Loss Cone Filling Rates}

For each galaxy we built simple spherical models using the same
Nuker law parameters as those in \citet{wang04} and separately
calculated the loss cone filling rates due to two-body interaction
in single and binary BH cases (see MT99 and Y02 for detailed
description of the calculation).

For single BHs we did two sets of calculations to investigate the
effect GR on loss cone filling rate $F^{\rm lc}_{\rm single}$. In the
non-GR case we calculated $J_{\rm lc}$ with $r_t$ from
equation~(\ref{rt}) while in the GR case we substituted $r_t$ with
$2r_{\rm S}$ if $r_t<2r_{\rm S}$.  In both cases the resulting loss
cone filling rates are similar, within a factor of $1.4$ even when
$2r_{\rm S}\gg r_t$.  This is because the loss cones of the most
massive BHs are diffusive so that $F^{\rm lc}_{\rm single}$ is
insensitive to the size of loss cone. In Table~\ref{table1} we only
give $F^{\rm lc}_{\rm single}$ for the non-GR case.  The mean of $\log
F^{\rm lc}_{\rm single}$ is $-4.82$ for all the galaxies or $-4.44$
for galaxies with $M_\bullet<10^8M_\odot$, which is within a factor of
$2$ of the rates obtained by \citet{wang04}, the differences being
completely accounted for by our different assumed BH masses.

For binary BHs GR effect is not important to loss cone filling rate
because we are interested in the case $a\gg r_S$. For each galaxy we
have calculated the loss cone filling rates for binary BHs with a
set of $a$ and derived the evolution timescale $t_{\rm evol}(a)$
with
\begin{equation}\label{tevol}
  \frac{1}{t_{\rm evol}(a)}=\frac{1}{t_h(a)}+\frac{1}{t_{\rm gr}(a)}
\end{equation}
(Y02), where $t_{\rm gr}$ is the evolution timescale due to
gravitational wave radiation \citep{peters64}. Then the binary's
stalling radius $a_{\rm stall}$ is determined by locating the peak
of the $t_{\rm evol}(a)$ curve. If $t_{\rm evol}(a_{\rm
stall})>10^{10}$ yr, which is the case for most of our model
galaxies, we increase $a_{\rm stall}$ until (1) $t_{\rm
evol}(a_{\rm stall})<10^{10}$ yr or (2) $a_{\rm stall}>0.1$
arcsec, whichever comes first. The latter criteria is motivated by
that SMBHBs in nearby galaxies, if wider than $0.1$ arcsec, would
have been observed. Notice that $a_{\rm stall}$, $t_{\rm
evol}(a_{\rm stall})$, and $F^{\rm lc}(a_{\rm stall})$ given by
either criteria (1) or (2) are independent of $q$. The results for
$q=0.1$, the typical mass ratio in the literature for the
present-day SMBHBs, are given in Table~\ref{table1}, while the
results for $q=0.01$ and $1$ are essentially identical. Although
the loss cone of SMBHB in $E_0-J^2$ space could be more than four
orders of magnitude wider than that of a single BH
(eq.s~[\ref{jlc}] and [\ref{rtah}]), the loss cone filling rate
for binary BH increases only by a factor of $10$, which implies
that the loss cone of SMBHB is also in the diffusive regime.

\subsection{Stellar Disruption Rates in the Non-relativistic Case}
\label{nongrdr}

First, we ignore GR effect and only consider tidal disruption of
solar type stars. In this case the stellar disruption rates for
single BHs have been given by column 4 in Table~\ref{table1}. For
binaries the rates are calculated according to equation~(\ref{nd})
with the loss cone filling rates listed in Table~\ref{table1} and
the non-relativistic cross sections from \S~\ref{nongrcs}.

\begin{figure}
  \plotone{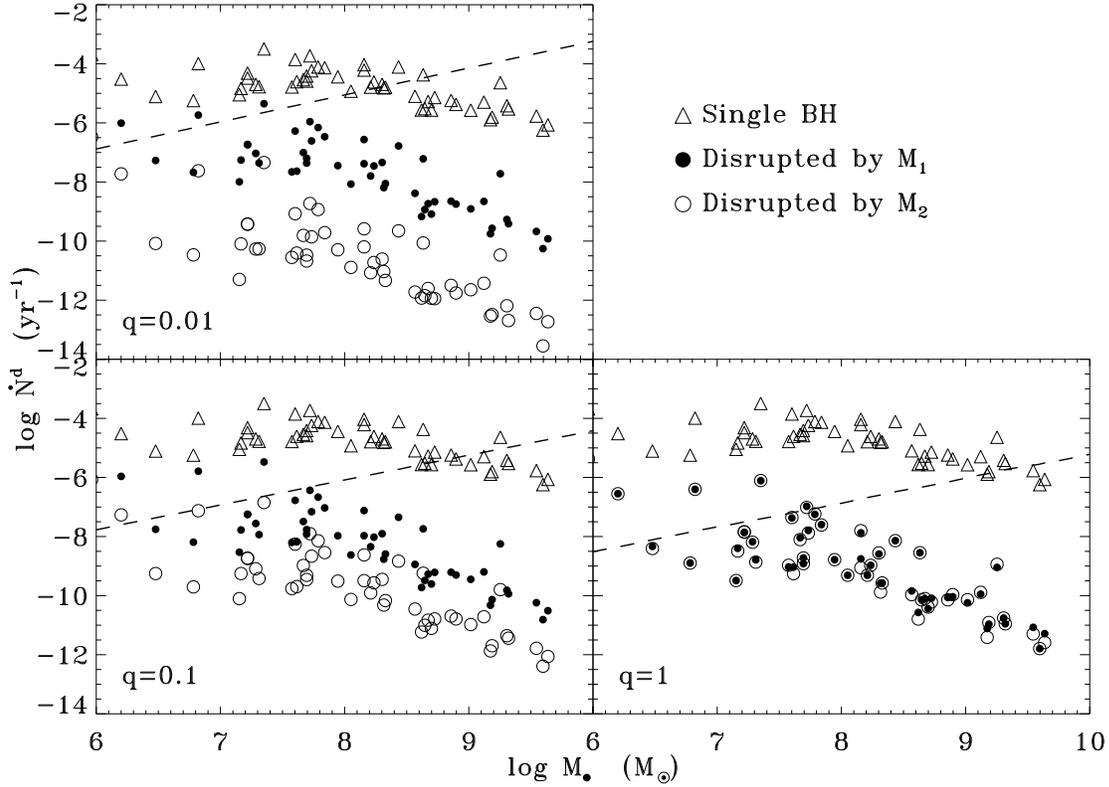}\\
\caption{Tidal disruption rates of solar-type stellar objects as a
function of total BH mass in Newtonian theory. Dots and circles,
respectively, refer to the disruption rates for primary and
secondary BHs. Triangle are for single BHs. Dashed lines are
thresholds above which SMBHBs would coalesce within $10^{10}$ yr.
The tidal disruption rates are derived under the assumptions that
each galaxy is spherical and that the loss cone refilling is due
to two-body relaxation.}\label{case1}
\end{figure}

In Figure~\ref{case1} we present the stelar disruption rates for
both single (triangles) and binary (dots and circles) BHs. It is
clear that the total disruption rates for SMBHB ($\dot{N}^d_{\rm
binary}=\dot{N}^d_1+\dot{N}^d_2$) is significantly lower than that
for single BH ($\dot{N}^d_{\rm single}$). The contrast between
$\dot{N}^d_{\rm binary}$ and $\dot{N}^d_{\rm single}$ is more than
one order of magnitude and increases with $q$. The significant
reduction of tidal disruption rates in binary BH systems originates
in the depletion of loss-cone stars and the suppression of loss cone
filling rate by the SMBHB.

Mechanisms other than two-body relaxation would increase the loss
cone filling rates in both single and binary SMBH systems (MT99;
Y02; \citealt{mm05}). The enhancement in loss cone filling rate is
uncertain because it depends on processes which are not well
understood. However, equation~(\ref{th}) implies that the
lifetime of hard SMBHB decreases with loss cone filling rate. So
for long-life SMBHB with $t_h\ga10^{10}$ yr the loss cone filling
rate should not significantly exceed $F^{\rm lc}_{\rm cri}\equiv
M_{12}/(2Km_*~10^{10}~{\rm yr})$. Correspondingly the stellar
disruption rate in SMBHB system is unlikely to be much greater than
\begin{eqnarray}
  \dot{N}^d_{\rm cri}\equiv F^{\rm lc}_{\rm
cri}\left[\frac{\Sigma_1(r_{t1})}{\Sigma_1(a_h)}+\frac{\Sigma_2(r_{t2})}{\Sigma_2(a_h)}\right]\,.\label{ndcri}
\end{eqnarray}
These upper limits of stellar disruption rates for binary BHs have
been presented in Figure~\ref{case1} as dashed lines. In $48$ out of
$51$ cases $\dot{N}^d_{\rm binary}$ is below $\dot{N}^d_{\rm cri}$,
reflecting that two-body relaxation is inefficient in refilling the
loss cone. For galaxies with $M_\bullet\la10^8M_\odot$
$\dot{N}^d_{\rm cri}$ is always lower than $\dot{N}^d_{\rm single}$,
so long-life SMBHBs in these galaxies would manifest themselves by
prominently suppressing stellar disruption rates no matter what
mechanism is responsible for the loss cone replenishing. At
$M_\bullet\ga10^8M_\odot$ $\dot{N}^d_{\rm cri}$ could be higher than
$\dot{N}^d_{\rm single}$ if $M_\bullet\gg10^8M_\odot$ or $q\ll1$,
implying that in these galaxies SMBHBs are not so distinct in the
efficiency of stellar disruption.

\subsection{Stellar Disruption Rates in the Relativistic
Case} \label{grdr}

In the relativistic case there is a critical mass $M_{\rm cri}$
above which a BH would swallow the whole star without tidal
disruption \citep{hills75}. For solar type stars recent
relativistic simulation \citep{ivanov06} gave that $M_{\rm
cri}\sim(4-8)\times10^7M_\odot$ if the BH is non-spinning and
$M_{\rm cri}\sim10^9M_\odot$ if the BH is maximally spinning. Due
to the ambiguity of BH spin it is not clear whether a star
approaching a BH more massive than $10^8M_\odot$ would necessarily
be disrupted and produce a flare. However, it is expected that
stellar disruptions by these massive BHs are rare because only
those stars approaching the spinning BH along a corotating orbit
close to the equatorial plane could get disrupted. For less
massive BHs ($M_\bullet\la 10^8M_\odot$) although
equation~(\ref{rt}) suggests that tidal disruption is inevitable
($r_t\ga r_{\rm S}$), part of the stellar debris with low angular
momentum may directly plunge into the event horizon without
producing a flare \citep{nolthenus83}. The proportion of such
debris is sensitive to the spin of the star just before tidal
disruption and is uncertain so far, but the proportion should be
low when $r_t\ll r_{\rm S}$. Because of all these uncertainties,
we adopted a rough assumption that flares are produced only in the
condition $r_t>r_{\rm S}$.

\begin{figure}
\plotone{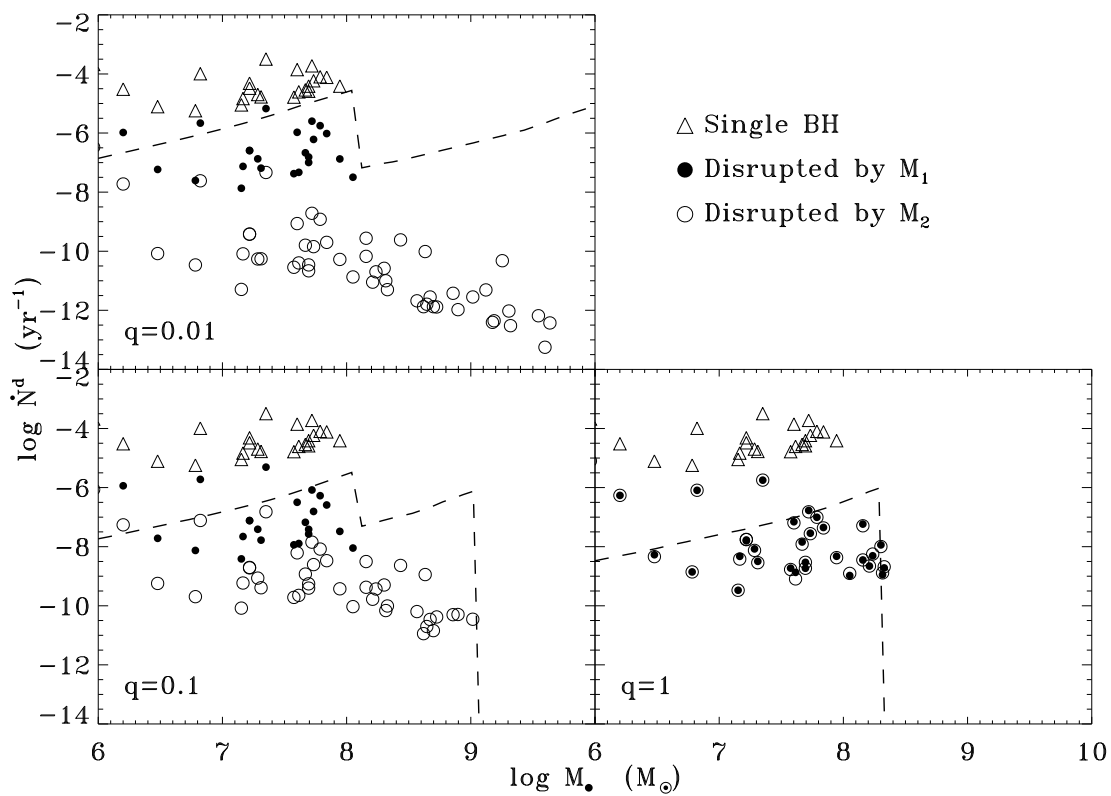}\\
\caption{Tidal disruption rates of solar-type stellar objects as a
  function of total BH mass, including general relativisitic effects.
Symbols and dashed lines have the same meanings as those in
Figure~\ref{case1}.}\label{flare}
\end{figure}

Figure~\ref{flare} shows the stellar disruption rates when GR effect
is taken into account. The tidal disruption rates for binary BHs are
calculated with the loss cone filling rates from Table~\ref{table1}
and the relativistic tidal disruption cross sections presented in
\S~\ref{grcs}. The resulting relativistic tidal disruption rates are
about $r_t/(r_t-r_{\rm S})$ times greater than the non-relativistic
ones if $r_t\geqslant2r_{\rm S}$, or $4r_{\rm S}/r_t$ times greater
if $r_{\rm S}<r_t<2r_{\rm S}$. The dashed lines show the thresholds
$\dot{N}^d_{\rm cri}$ above which SMBHBs would coalesce within
$10^{10}$ yr. They are derived according to equation~(\ref{ndcri})
but with relativistic cross sections. To study the frequency of
tidal flares we have set the stellar disruption rates zero when
$r_t<r_{\rm S}$, thus the sudden cutoff at $\sim(1+q)10^8M_\odot$
and $\sim q^{-1}(1+q)10^8M_\odot$ is artificial.

At $M_\bullet\ll10^8M_\odot$ the stellar disruption rates are
similar to those presented in Figure~\ref{case1} and again we find
that the stellar disruption rates for SMBHBs are considerably
lower than those for single BHs. While at
$M_\bullet\gg10^8M_\odot$ the only sources producing tidal flares
are the secondary BHs, though the rates are low.

\subsection{Effect of a Spectrum of Stellar Masses}

The results of previous sections are based on the assumption that
each galaxy is composed of stars with solar mass and radius. Since
none of our sample galaxies exhibits recent nuclear star formation,
taking into account a mass spectrum of main-sequence stars would
result in numerous low mass stellar objects which have smaller mean
tidal disruption cross sections and are less efficient to relax via
two-body interaction. But these effects would not qualitatively
change the tidal disruption rates since: (a) the tidal radius is not
sensitive to the stellar mass because $r_*\sim m_*^{0.47}$
\citep{bond84} so that $r_t\sim m_*^{0.14}$; (b) the increment in
stellar number density to reproduce the mass distribution inferred
from observations compensates for the decrease in the efficiency of
two-body relaxation.

However, tidal disruption of off-main-sequence giant stars becomes
important when BHs are more massive than $10^8M_\odot$ because in
this case the tidal radius is greater than the Schwarzschild radius.
Following MT99 we assume that giant stars with time-averaged radius
$r_g=15r_\odot$ contribute $g=1\%$ to the total stellar population.
These values are consistent with the approximate stellar evolution
model given in \citealt{syer99}. For single BHs the disruption rates
of giants in the diffusive loss cone limit are obtained with
$gF^{\rm lc}_{\rm single}$ (MT99), where the loss cone filling rates
$F^{\rm lc}_{\rm single}$ in relativistic case are from
Table~\ref{table1}. For binary BHs the disruption rates are
calculated in the same way as in \S~\ref{grdr} except that the loss
cone  filling rates and tidal radii for giant stars are used. Since
we are interested the frequency of tidal flares we have set the
stellar disruption rates zero when $r_t<r_{\rm S}$.

\begin{figure}
\plotone{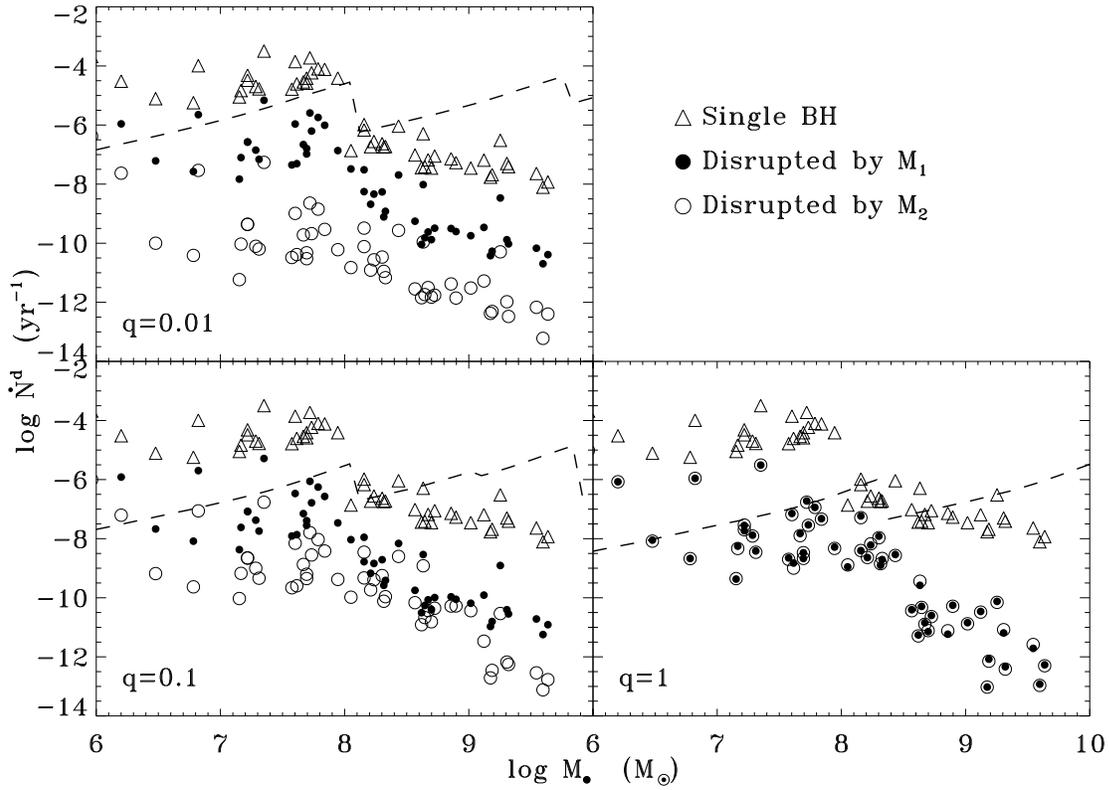}\\
\caption{Total tidal disruption rates both of stellar-type stellar
  objects and red giants as a function of total BH mass, including
  general relativistic effect and
assuming that red giants contribute $1\%$ to the loss cone filling
rate. Symbols and dashed lines have the same meanings as those in
Figure~\ref{case1}.}\label{giant}
\end{figure}

Figure~\ref{giant} shows the stellar disruption rates and the
thresholds $\dot{N}^d_{\rm cri}$ when both solar type and giant
stars are taken into account. At $M_\bullet\la10^8M_\odot$ the
stellar disruption rates are dominated by disruption of solar type
stars therefore similar to those in Figure~\ref{flare}. At
$M_\bullet>10^8M_\odot$ disruption of giant stars take over in
both single and binary BH systems, resulting in low but no longer
zero flaring rates. The suppression of stellar disruption rate by
SMBHB is still obvious if the loss cone refilling is dominated by
two-body relaxation, but other mechanisms could potentially
enhance the stellar disruption rates for massive SMBHBs
($M_\bullet>10^8M_\odot$) to a level indistinguishable from those
for single BHs.

\section{RATES OF TIDAL FLARES IN LOCAL UNIVERSE}\label{prediction}

\begin{figure}
\plotone{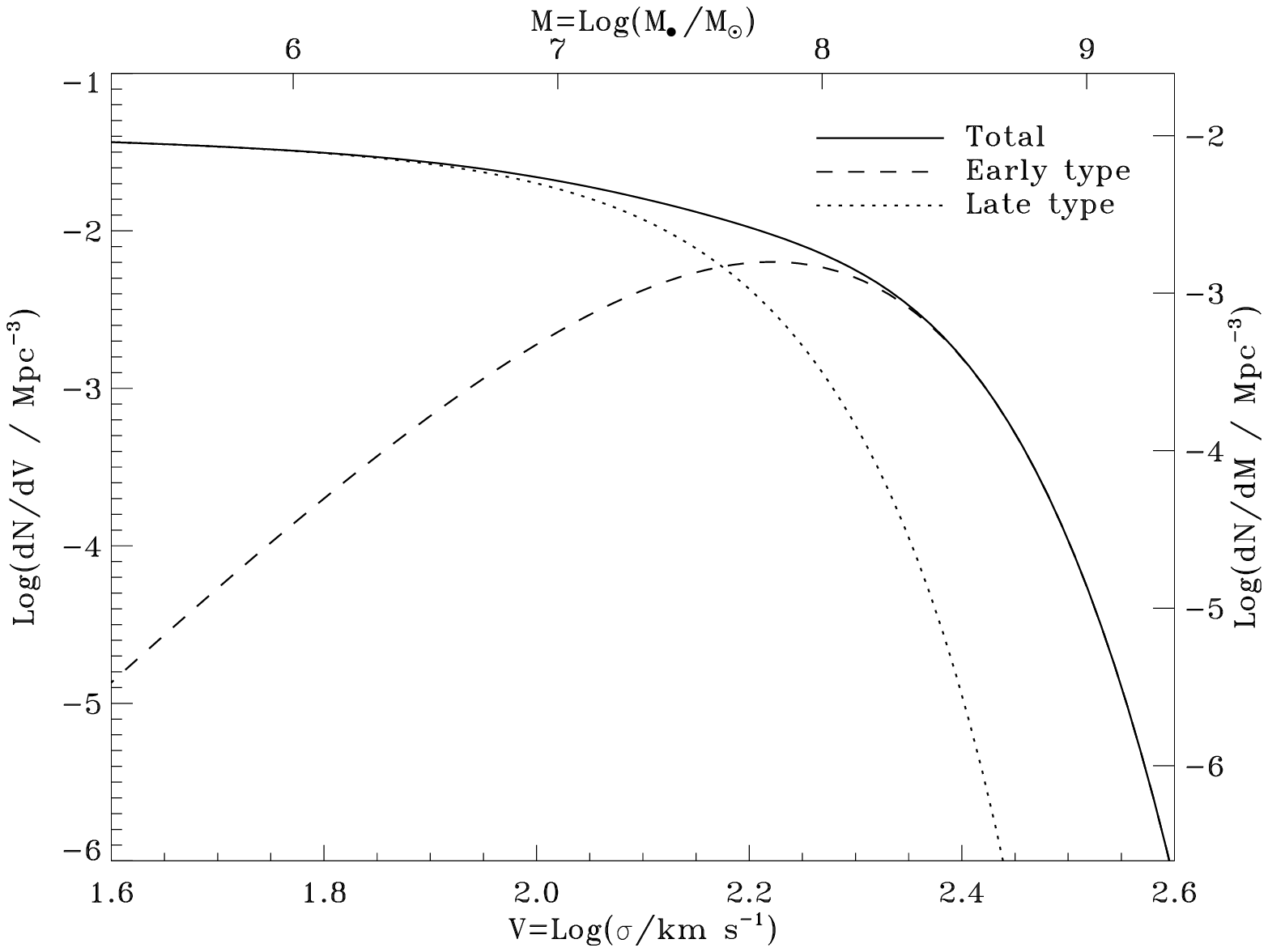}\\
\caption{number density of SMBHs as a function of stellar velocity
dispersion or BH mass \citep{marconi04}.}\label{mf}
\end{figure}

To estimate the density of tidal flares in local universe we adopted
the mass functions of SMBH for early (E+S0) and late (Sabcd) type
galaxies from \citet{marconi04}, which are converted from the
distribution of stellar velocity dispersion using the empirical
$M_\bullet-\sigma_*$ relation \citep{tremaine02}. The mass functions
are presented in Figure~\ref{mf} and in the following we assume that
the total BH mass ($M_{12}$) in SMBHB systems follows the same
distribution. At $M_\bullet<10^6M_\odot$ the mass distribution of
BHs is unclear because the demographics of the BHs in dwarf
galaxies is not well established. The total BH mass density
according to Figure~\ref{mf} is $2.5\times10^5~M_{\odot}~{\rm
Mpc^{-3}}$ ($H_0=70~{\rm km~s^{-1}~Mpc^{-3}}$ throughout this
paper), consistent with the value given by \citet{yu02a}.

The flaring rates per unit volume are calculated by integrating the
BH mass functions weighted by the flaring rates. The flaring rates
for single and binary BHs are from Figures~\ref{flare} (we have
smoothed the flaring rates) and the fiducial field of integration is
$M_\bullet(M_{12})\in[10^6M_\odot,10^{10}M_\odot]$. In single BH
case we do a second set of calculations with
$M_\bullet\in[10^3M_\odot,10^{10}M_\odot]$ to account for the
flaring rates from the intermediate-mass BHs (IMBHs,
$10^3M_\odot<M_\bullet<10^6M_\odot$) at the centers of dwarf
galaxies, while for binary BHs we skip such calculations because
most dwarf galaxies have not experienced mergers \citep{haehnelt02}.
The flaring rates at $M_\bullet\leqslant10^6M_\odot$ are from the
linear extrapolation of the logarithmic flaring rate at
$M_\bullet>10^6M_\odot$. We do not consider flares produced by
giants because we will see below that these flares differ from those
produced by normal stars. Our result is not sensitive to this
treatment because the mass function of SMBHs cuts off steeply  above
$10^8M_\odot$ where flares from giants become dominant.

\begin{deluxetable}{llllll}
    \tablewidth{0pt}
    \tablecaption{Flaring rate per unit volume in ${\rm yr^{-1}~Mpc^{-3}}$ \label{table2}}
    \tablehead{
    \colhead{Galaxy type} &
    \colhead{$\dot{N}^f_{\rm single}$} &
    \colhead{$\dot{\cal N}^f_{\rm single}$} &
    \colhead{$q$} &
    \colhead{$\dot{N}^f_1$} &
    \colhead{$\dot{N}^f_2$}\\
    \colhead{(1)} &
    \colhead{(2)} &
    \colhead{(3)} &
    \colhead{(4)} &
    \colhead{(5)} &
    \colhead{(6)}
    }
    \startdata
    Early & $6.0\times10^{-8}$ & $6.0\times10^{-8}$ & 0.01 & $3.6\times10^{-10}$ & $3.8\times10^{-13}$ \\
    Early & $6.0\times10^{-8}$ & $6.0\times10^{-8}$ & 0.1  & $1.2\times10^{-10}$ & $2.4\times10^{-12}$ \\
    Early & $6.0\times10^{-8}$ & $6.0\times10^{-8}$ & 1    & $2.1\times10^{-11}$ & $2.1\times10^{-11}$ \\
    Late  & $3.0\times10^{-7}$ & $1.4\times10^{-6}$ & 0.01 & $3.8\times10^{-9}$  & $2.2\times10^{-11}$ \\
    Late  & $3.0\times10^{-7}$ & $1.4\times10^{-6}$ & 0.1  & $2.3\times10^{-9}$  & $9.0\times10^{-11}$ \\
    Late  & $3.0\times10^{-7}$ & $1.4\times10^{-6}$ & 1    & $5.6\times10^{-10}$ & $5.6\times10^{-10}$ \\
    \enddata

\tablecomments{Column 1 gives the galaxy type: 'Early'$=$E$+$S0,
'Late'$=$Sabcd. Columns 2 and 3 are the flaring rates per unit
volume in single BH case, the former being calculated for
$M_\bullet>10^6M_\odot$ while the latter for
$M_\bullet>10^3M_\odot$. $q$ is the assumed mass ratio of SMBHB,
and $\dot{N}^f_1$ and $\dot{N}^f_2$ are respectively the flaring
rates for primary and secondary BHs in binary BH case.}

\end{deluxetable}

The flaring rates per unit volume for different types of galaxies
and for both single and binary BHs are presented in
Table~\ref{table2}. Note that taking into account asymmetric stellar
distributions and relaxation mechanisms other than two-body
interaction would significantly increase the loss cone filling rates
in both single and binary BH systems, therefore dramatically
increase the flaring rates. According to Table~\ref{table2}, it is
obvious that if SMBHBs are ubiquitous in the centers of both early
and late type galaxies the flaring rate in local universe would be
more than one order of magnitude lower than that in single BH case.
It is also clear that dwarf galaxies, if they follow the same mass
distribution as that of massive galaxies and harbor IMBHs at their
centers, would contribute the majority to the flares in late type
galaxies but have little effect on the flaring rate in early type
galaxies.

\begin{figure}
\plotone{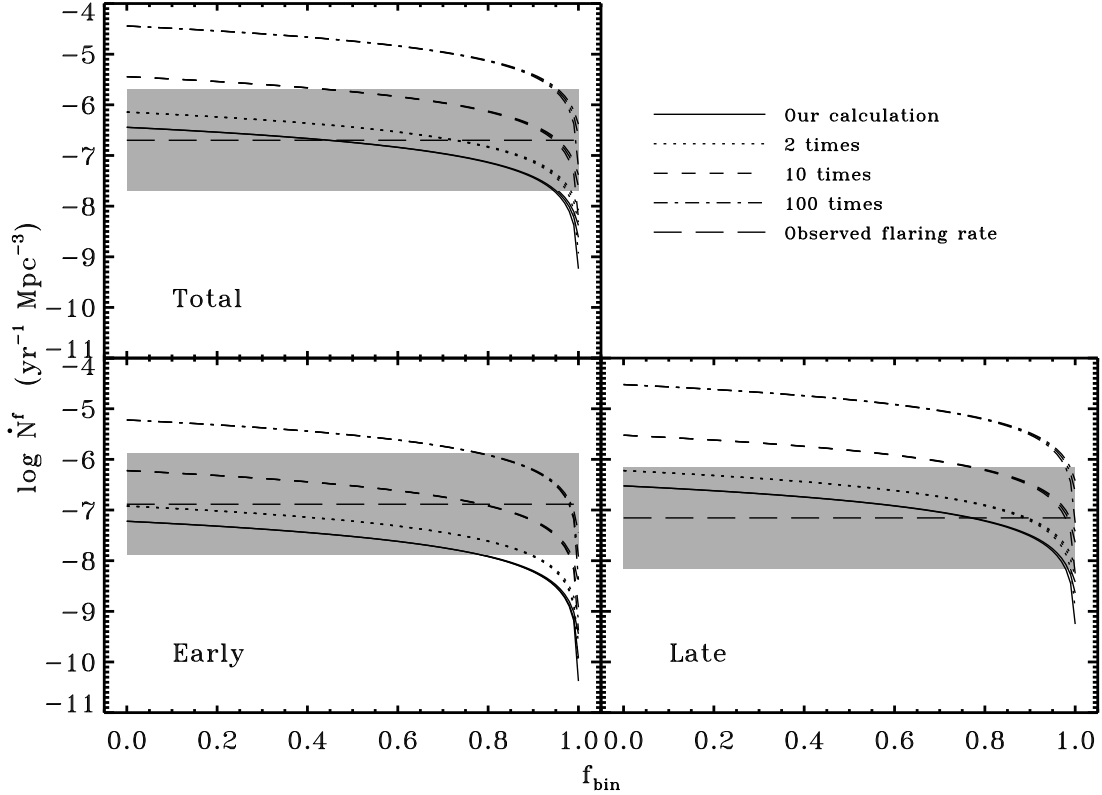}\\
\caption{Expected tidal disruption rates of stellar objects by
black  holes as a function of the fraction of binaries for all
(upper), early (lower left), and late (lower right) type galaxies.
Solid lines are our calculated flaring rates for spherical
galaxies with two-body relaxation. Dotted, short-dashed, and
dot-dashed lines are calculated by simply multiplying the solid
lines with, respectively, $2$, $10$, and $100$. The long-dashed
horizontal lines indicates the best estimates of the flaring rates
in real universe and the shaded regions corresponding to an
assumed statistical uncertainty of one order
magnitude.}\label{constrain}
\end{figure}

If a fraction, $f_{\rm bin}$, of SMBHs in the investigated
galaxies are in binary and a fraction $1-f_{\rm bin}$ are single,
the expected tidal disruption rate is
\begin{eqnarray}
  \dot{N}^f &=& (1-f_{\rm bin})\dot{N}^f_{\rm single}+f_{\rm
  bin}(\dot{N}^f_{1}+\dot{N}^f_{2})\,.
\end{eqnarray}
$\dot{N}^f$ as a function of $f_{\rm bin}$ is presented in
Figure~\ref{constrain}. It shows that $\dot{N}^f$ is not sensitive
to $q$ unless $f_{\rm bin}\simeq1$. In \S~\ref{discussion} we will
use these $f_{\rm bin}-\dot{N}^f$ diagrams to constrain $f_{\rm
bin}$ for different types of galaxies.

\section{DISCUSSION}\label{discussion}

The hierarchical structure formation model in CDM cosmology
predicts that SMBHBs are continuously formed across the merging
history of galaxies. Constraining the abundance of SMBHBs among
inactive galaxies is essential to test the theoretical models of
the dynamic evolution of binary BHs. In this paper we have studied
the effect of hard binary BHs on stellar disruption rates, trying
to find distinct observational signatures of SMBHBs in galaxy
centers. We focus our attention on inactive galaxies which contain
the final products of BH mergers so we do not consider the effect
of gas on SMBHB evolution.

We have carried out numerical scattering experiments designed for
hard SMBHBs to investigate the probability of stellar disruption.
Since $r_t\ll a_h$, in each set of experiments usually a large
number of test particles ($N\simeq10^7$) were used to give
statistical meaningful results. For solar type stars the test
particle approximation should be valid outside $r_t$ as long as
$E_0\ll GM_{12}/a$. In this situation the self-binding energy of a
star $\sim Gm_*^2/r_*$ is smaller by a factor
$(M_\bullet/M_*)^{2/3}$ than its kinetic energy at $r_t$ so the
loss of orbital energy due to tidal dissipation is not important.
The test particle approximation may be problematic for giant stars
because they are much less concentrated and may suffer from mass
stripping during their encounters with BHs. This uncertainty could
be incorporated into the uncertainty in $r_g$. In single BH case
the stellar disruption rate is not sensitive to $r_g$ if the loss
cone is in the diffusive regime but would be proportional to $r_g$
if the loss cone is in the pinhole regime (MT99). For binary BHs
the loss cone filling rate does not depend on $r_g$ but the tidal
disruption cross section scales with $r_g$ when gravitational
focusing is important. Therefore unless the effective $r_g$
deviates significantly from our choice $r_g=15r_\odot$ our results
for giant stars would not be qualitatively different.

We find that in binary BH systems the multi-encounter cross sections
are greater than the single-encounter ones. This is because binary
BHs tend to scatter particles onto temporary bound orbits which
enhances the number of star-binary encounters. If a star is only
partly disrupted during each close passage from the BHs,
multi-flares would be produced occur due to the multi-encounters,
which provides a distinct observational signature of SMBHBs.
However, we have seen in \S~\ref{nongrcs} that the probability of
multi-flaring may be extremely low because $r_{t}\ll a$. Even if
$r_{t}\sim a$ equation~(\ref{rsrt}) implies that $a/r_{{\rm S}1}\sim
M_8^{-2/3}$ so gravitational wave radiation would quickly drive the
BHs to coalesce, making the multi-flares unlikely. A secondary flare
could also be produced if the secondary member of SMBHB captures
sufficient debris from the star disrupted by the primary BH, or vise
versa. The probability of these events depends on the opening angle
of all the spewed debris about the stellar disrupting BH, which
needs to be determined by hydrodynamic simulations.

We have shown that the normalized cross sections are not sensitive
to the star's initial binding energy or to the eccentricity of the BH
binary, but do depend on the amplitude and orientation of the star's
initial angular momentum. The cross sections presented in
Figures~\ref{cs001} and \ref{fit}, which have been equally averaged
over the whole loss cone and over all directions, can be directly
used if the loss cone of SMBHB is isotropic and is in the pinhole
regime. In the diffusive loss cone limit the stars falling to the binary
BHs satisfy $J\ga J_{\rm lc}$. According to Figure~\ref{dfjdr} the
differential tidal disruption cross sections at $J\ga J_{\rm lc}$
are lower than those inside the loss cone, so the mean tidal
disruption cross section in the diffusive limit is smaller
than those presented in Figures~\ref{cs001} and \ref{fit}.

We find that the tidal disruption rates in binary BH systems are
more than one order of magnitude lower than those in single BH
systems. The suppression of stellar disruption rate by SMBHB
originates in the low loss cone filling rate in binary BH systems,
which results from the stellar cusp destruction due to the
hardening of SMBHBs. Even efficient loss cone filling mechanisms
are taken into account the contrast between the stellar disruption
rates in single and binary BH cases are still prominent in less
massive galaxies with $M_\bullet\la10^8M_\odot$, because the
survival of SMBHBs requires that the loss cone filling rate should
not exceed a critical value which is proportional to the total BH
mass of SMBHB. Our result is qualitatively different from that of
\citet{ivanov05}, who suggested that a SMBHB will enhance the
stellar disruption rates. The difference is due to that we are
studying a much later evolution stage at which the dense galactic
cusps considered by \citeauthor{ivanov05} have been destroyed
during the hardening of SMBHBs. According to the calculation in
Y02 binary BHs reach the hard radius in about a dynamic friction
timescale of the merging galaxy. So the situation considered by
\citet{ivanov05} applies for a short period relative to the
lifetime of a galaxy. \citet{merritt05} suggested that because
loss cone refilling takes time the stellar disruption rate
immediately after SMBHB coalescence is significantly below the
final steady disruption rate. But this effect is prominent only in
massive galaxies with $M_\bullet>10^8M_\odot$, where the
relaxation timescale is long. Our result implies that in less
massive galaxies the flaring rate could give stringent constraint
to the abundance of SMBHB.

In larger galaxies with $M_\bullet>10^8M_\odot$ tidal flares are
dominated by disruption of giants. However, their spectral and
variable properties may be different from those of solar type stars.
A tidal flare is expected to be produced at a scale about $r_t$,
initially radiating at the Eddington limit $L_{\rm
Edd}\simeq1.3\times10^{45}(M_\bullet/10^7M_\odot)~{\rm erg~s^{-1}}$
with a thermal spectrum of effective temperature $T_{\rm
eff}\simeq[L_{\rm Edd}/(4\pi
r_t^2\sigma)]^{1/4}=2.4\times10^5(r_*/r_\odot)^{-1/2}(m_*/M_{\odot})^{1/6}M_8^{1/12}~{\rm
K}$ and decaying on a timescale $t_{\rm flare}\sim \pi
GM_{\bullet}(2\Delta E)^{-3/2}\simeq 1.1~{\rm
yr}(r_*/r_\odot)^{3/2}(m_*/M_\odot)^{-1}M_8^{1/2}$, where $\Delta
E\sim GM_\bullet r_*/r_t^2$ characterizes the span of the binding
energy of the debris \citep{rees88}. For solar type stars the
spectrum of tidal flare peaks at UV or soft X-ray and the luminosity
decays in about $M_8^{1/2}$ year. While for giant stars with
$r_*=15r_\odot$ and $m_*=m_\odot$ the spectrum peaks at optical or
UV band and the flare dims on a timescale of about $60M_8^{1/2}$
years. Therefore if a UV/X-ray outburst decaying within months to
one year is detected in a galaxy with total BH mass much greater
than $10^8M_\odot$ (determined by other observations), it is likely
that a less massive secondary SMBH resides in the galactic center.
Such event is another signature of SMBHB, though Figure~\ref{giant}
suggests that the frequency of such event is low.

So far six candidate tidal disruption events have been observed in
nearby inactive galaxies. Among them, one was observed at $z=0.37$
by the UV telescope \textit{GALEX} \citep{gezari06} and the rest
were discovered in galaxies at $z\la0.15$ during the one year
ROSAT All Sky Survey (\textit{RASS}) \citep{komossa02}. The Hubble
types for the host galaxies of the five \textit{RASS} flares are
not well determined: NGC 5905 is of a SB galaxy and the other four
galaxies look like ellipticals \citep{komossa02}. From the
observations of \textit{RASS} \citet{donley02} inferred a total
flaring rate of $\dot{N}^f_{\rm obs}\sim2\times10^{-7}~{\rm
yr^{-1}~Mpc^{-3}}$ for all types of galaxies. From the {\it RASS}
results we can also give a rough estimate of the flaring rates for
different types of galaxies: $\dot{N}^f_{\rm
obs}\sim1.3\times10^{-7}~{\rm yr^{-1}~Mpc^{-3}}$ for early type
galaxies and $\dot{N}^f_{\rm obs}\sim7\times10^{-8}~{\rm
yr^{-1}~Mpc^{-3}}$ for late types. We plot $\dot{N}^f_{\rm obs}$
as long-dashed horizontal lines in Figure~\ref{constrain} and the
corresponding uncertainties with shaded areas. Because of the poor
statistics in $\dot{N}^f_{\rm obs}$ and the ill-determination of
Hubble types of the host galaxies, the results are highly
uncertain.

We did the calculation based on the assumptions that galaxies are
spherical and two-body scattering relaxation dominates the
loss-cone refilling. However, taking into account more realistic
non-spherical stellar distributions and other loss-cone refilling
mechanisms besides two-body interaction will probably significantly
increase the loss-cone refilling rates in both single and binary
SMBH systems. MT99 and Y02 showed that taking into account
axisymmetric stellar distribution will increase the loss cone
refilling rates by a factor of $2$ for both single SMBHs and hard
SMBHBs. The loss cone refilling rates in triaxial galaxies are
slightly higher than those in axisymmetric ones if the triaxiality
is less than $0.1$ but can be two orders of magnitude higher if
the triaxiality is more extreme \citep[Y02;][]{merritt04}.
Resonant relaxation would increase the loss cone refilling rates
by a factor of $2-10$ relative to those in the case of two-body
relaxation, depending on the eccentricities of stellar orbits
\citep{rauch96,rauch98,gurkan07}. To calculate the tidal
disruption rates by taking into account the above physical
processes is out of the scope of this paper. To illustrate the
effects of increased loss-cone refilling rates and tidal
disruption rates on the estimated fraction of binary BHs, $f_{\rm
bin}$, we simply multiply our calculated stellar disruption rates
$\dot{N}^f$ by $2$, $10$, and $100$ and plot the results in
Figure~\ref{constrain}. In Figure~\ref{constrain}, the
intersections of long-dashed and solid lines give the lower limit
to $f_{\rm bin}$, while the intersections of long-dashed and
dot-dashed lines roughly give the upper limits. When all types of
galaxies are considered, binary fraction $f_{\rm bin}$ is between
$0.4$ and $1.0$. For early type galaxies, the solid lines do not
intersect the long-dashed line, implying that other loss-cone
filling mechanisms in addition to two-body relaxation should also
be important. For late type galaxies, Figure~\ref{constrain}
suggests a binary fraction $f_{\rm bin}\sim75\%$ and that extreme
triaxiality may not be common, otherwise the current observation
would imply an extremely high fraction of SMBHBs. We would like to
note that the constraints on $f_{\rm bin}$ are very uncertain due
to the large uncertainty in the current $\dot{N}^f_{\rm obs}$.
Detection of more candidate tidal disruption events is needed to
give better constraints on $f_{\rm bin}$ and on the dominant
loss-cone filling mechanisms in different types of galaxies.

The non-detection of X-ray or UV flares in dwarf galaxies seems
inconsistent with the calculated high flaring rates. The
discrepancy may be due to (1) a lack of sufficient sensitivity in
current surveys, (2) the continuous and steady accretion of
stellar debris onto IMBHs during successive tidal disruption
events \citep{milosavljevic06}, or (3) a lack of IMBHs in the
centers of dwarf galaxies. Future X-ray and UV surveys with
improved sensitivity and angular resolution would help
constraining the fraction of IMBHs in galaxy centers and give
interesting implications to the formation and the merging history
of IMBHs \citep{volonteri03,volonteri07}.

\section{CONCLUSIONS}\label{conclusion}

To investigate whether SMBHBs are ubiquitous in nearby inactive
galaxies or coalesce rapidly in galaxy mergers, we have studied the
tidal disruption rates of stellar objects in both single and binary
SMBH systems. We have calculated the interaction cross sections
between hard SMBHBs and intruding stars by carrying out intensive
numerical scattering experiments with typically $10^7$ particles,
taking into account the initial binding energy and angular momentum of
particle, the eccentricity of the orbit of SMBHB, and including
general relativistic effects.  We have also derived empirical formulae
for the relativistic cross sections, which can be applied to SMBHBs
with a wide range of semimajor axis and todal BH mass.

We have calculated the rate of loss cone refilling due to two-body
relaxation for a sample of $51$ nearby galaxies, assuming that each
galaxy is spherical. The steady loss cone filling rates in binary BH
systems would be significantly suppressed due to the three-body
interaction between SMBHBs and stars passing by. We have calculated
the tidal disruption rates respectively for single and binary SMBHs
by combining the loss cone filling rates and the tidal disruption
cross sections. We find that the tidal disruption rate in SMBHB
systems is more than one order of magnitude lower than that in
single SMBH systems. For galaxies with BHs more massive than
$10^8M_\odot$, a UV/X-ray flare at galactic center decaying within
one year provides strong evidence of a secondary BH, although the
probability of such events is low.

Finally we have calculated the flaring rates in local universe using the
BH mass function given in the literature. The comparison of the
calculated flaring rates and the preliminary results from current
X-ray surveys could not yet tell whether SMBHBs are ubiquitous in
local universe. Future UV/X-ray surveys with improved
sensitivity and duration are needed.

\acknowledgments

We are grateful to Dr. Xue-Bing Wu, Dr. Stefanie Komossa, and Dr.
Vladimir Karas for many constructive comments. We thank the
referee for helpful comments and suggestions. Many thanks are due to
Bing-Xiao Xu, Ran Wang, Lei 
Qian and Zhao-Yu Li for fruitful discussions. During this work we
have used the SGI Altix 330 system at the Department of Astronomy,
Peking University (PKU) and CCSE-I HP Cluster of PKU. This work is
supported by the National Natural Science Foundation of China (No.
10573001) and by the national 973 program (No. 2007CB815405). JM
thanks the Royal Society for financial support.

\end{document}